\documentclass{aa}

\input psfig.tex
\usepackage{graphicx}
\usepackage{natbib}
\usepackage{array}
\usepackage{graphics}
\usepackage{latexsym}
\usepackage{amssymb}
\usepackage{amsmath}
\usepackage{fancyhdr}
\usepackage{morefloats}
\usepackage{bm}
\usepackage{graphicx}
\usepackage{txfonts}
\begin{document}
\title{Thick disk kinematics from RAVE and the solar motion}

\author{S. Pasetto\inst{1,2},
E.K. Grebel\inst{2}, 
T. Zwitter\inst{3,4}, 
C. Chiosi\inst{5}, 
G. Bertelli\inst{6},
O. Bienayme\inst{7},
G. Seabroke\inst{1},
J. Bland-Hawthorn\inst{8},
C. Boeche\inst{2}, 
B.K. Gibson\inst{9,10},
G. Gilmore\inst{11},
U. Munari\inst{6}, 
J.F. Navarro\inst{12},
Q. Parker\inst{13,14,15},
W. Reid\inst{13,14},
A. Silviero\inst{5,16},
M. Steinmetz\inst{16}}

\offprints{sp2@mssl.ucl.ac.uk}

\institute{ %1
University College London, Department of Space \& Climate Physics, Mullard Space Science Laboratory, Holmbury St. Mary, Dorking Surrey RH5 6NT, United Kingdom
\and %2
Astronomisches Rechen-Institut, Zentrum f\"ur Astronomie der Universit\"at Heidelberg, M\"onchhofstr. 12-14, 69120 Heidelberg, Germany
\and %3
University of Ljubljana, Faculty of Mathematics and Physics, 1000, Ljubljana, Slovenia
\and %4
Center of Excellence SPACE-SI, A\v{s}ker\v{c}eva cesta 12, 1000, Ljubljana, Slovenia
\and %5
Department of Physics and Astronomy ``G. Galilei'', Padova University, Vicolo dell'Osservatorio 3, 35122 Padova, Italy
\and %6
INAF - Padova Astronomical Observatory, Vicolo dell'Osservatorio 5, 35122 Padova, Italy 
\and %7
Observatoire astronomique de Strasbourg 11 rue de l'Université, 67000 Strasbourg, France
\and %8
Sydney Institute for Astronomy, University of Sydney, NSW 2006, Australia
\and %9
Jeremiah Horrocks Institute, University of Central Lancashire, Preston, PR1 2HE, UK 
\and %10
Monash Centre for Astrophysics, Monash University, Clayton, 3800, Australia
\and %11
Institute of Astronomy, Cambridge University, Madingley Road, Cambridge CB3 0HA
\and %12
University of Victoria, Department of Physics and Astronomy, Victoria, BC Canada V8P 5C2
\and %13
Department of Physics and Astronomy, Macquarie University, NSW 2109, Australia
\and %14
Macquarie research centre in Astronomy, Astrophysics and Astrophotonics, Macquarie University NSW 2109, Australia
\and %15
Australian Astronomical Observatory, PO Box 296, Epping, NSW 2121, Australia
\and %16
Leibniz-Institut fur Astrophysik Potsdam (AIP), An der Sternwarte 16, 14482 Potsdam, Germany
}
\date{Accepted for publication on A\&A}

\titlerunning{Thick disk with RAVE}
\authorrunning{S.\ Pasetto et al.}

\abstract
{Radial velocity surveys such as the Radial Velocity Experiment (RAVE) provide us with measurements of hundreds of thousands of nearby stars most of which belong to the Galactic thin, thick disk or halo. Ideally, to study the Galactic  disks (both thin and thick) one should make use of the  multi-dimensional phase-space and the whole pattern of chemical abundances of their stellar populations. }
{In this paper, with the aid of the RAVE Survey,  we study the thin and thick  disks of the  Milky Way, focusing on the latter. We present a technique to disentangle the stellar content of the two disks based on the kinematics and other stellar parameters such as the surface gravity of the stars. Using the Padova Galaxy Model, we checked the ability of our method to correctly isolate  the thick disk component from  the Galaxy mixture of stellar populations.}
{We introduce selection criteria in order to clean the observed radial velocities from the Galactic differential rotation and to take into account the partial sky coverage of RAVE. We developed a numerical technique to statistically disentangle thin and thick disks from their mixture.}
{We deduce the components of the solar motion relative to the Local Standard of Rest (LSR) in the radial and vertical direction, $(v_U , v_W)_{\odot} = \left( {9.87 \pm 0.37,8.01 \pm 0.29} \right){\text{km s}}^{ - 1}$,
 the rotational lag of the thick disk component relative to the LSR $	 {{v}_{\text{thick,lag}}}\cong 49{\mbox{ }\text{km s}}^{-1}$, and the square root of the absolute value of the velocity dispersion tensor for the thick disk alone: $\sigma _{RR}^{}  = (56.1 \pm 3.8) {\text{km s}}^{ - 1}$, $\sigma _{R\varphi }  = (29.4 \pm 17.2) {\text{km s}}^{ - 1}$, $\sigma _{Rz}  = (10.1 \pm 3.3) {\text{km s}}^{ - 1}$, $\sigma _{\varphi \varphi }  = (46.1 \pm 6.7) {\text{km s}}^{ - 1}$, $\sigma _{\varphi z}  = (5.8 \pm 5.1) {\text{km s}}^{ - 1}$, $\sigma _{zz}  = (35.1 \pm 3.4) {\text{km s}}^{ - 1}$. The analysis of the thin disk is presented in another paper. We find good agreement with previous independent parameter determinations. In our analysis we used photometrically determined distances. In the Appendix we show that similar values can be found for the thick disk alone as derived in the main sections of our paper even without the knowledge of photometric distances.
}
{}


\maketitle

\section{Introduction}\label{Introduction}

Stellar radial velocity surveys such as the Radial Velocity Experiment RAVE
\citep{2006AJ....132.1645S,2008AJ....136..421Z,2011AJ....141..187S} can provide, for all the observed stars, three of the six components
of the Galactic stellar phase space - two directional components and one velocity component. The lack of the other three components makes it difficult to fully
characterize the dynamical behaviour of our Galaxy and its present
phase-space description from such surveys.

Despite these limitations, RAVE has the unique merit of systematically measuring stellar radial velocities with a precision never previously realized for such a large sample of stars,
providing us with a statistically significant amount of data that continues
to grow, recently surpassing the 500,000 spectra. RAVE is a southern-hemisphere survey obtaining spectra in the near-infrared CaII triplet region of magnitude selected stars ($9<I<13$). The usefulness of this vast dataset has already been
demonstrated in the context of kinematics in order to deduce the
characteristics of the stellar velocity distributions \citep[e.g., ][]{2008A&A...480..753V, 2011ApJ...728....7C, 2008MNRAS.391..793S, 2011MNRAS.418.2459H, 2011MNRAS.412.1237C, 2011MNRAS.412.2026S, 2011MNRAS.413.2235W, 2012NewA...17...22K, 2012MNRAS.421.3362B}, to identify the  stellar streams \citep[e.g.,][]{2008MNRAS.384...11S,2008ApJ...685..261K, 2011ApJ...728..102W, 2011MNRAS.411..117K}, and to study the chemistry of MW components \citep[e.g.,][]{2010ApJ...721L..92R, 2011ApJ...743..107R, 2010ApJ...724L.104F, 2011AJ....142..193B, 2012MNRAS.419.2844C}. Here, we carry out an analysis on topics recently addressed by \citet{2011MNRAS.412.1237C} and  \citet{2011ApJ...728....7C}, leaving aside for the moment the interpretative tools based on the orbit integration as developed by \cite{2011MNRAS.413.2235W}. In other words, although we will make use of a Galactic potential model, the evolution with time $t$ will not be the direct object of our analysis.
The magnitude range of RAVE ($9 < I < 13$) \citep[see e.g.,][]{2006AJ....132.1645S,
2008AJ....136..421Z, 2011MNRAS.412.2026S} implies that most of the stars targeted by RAVE
belong to the thin or the thick disk of the MW.  Moreover, working with
RAVE data to investigate the kinematics of our Galaxy in a statistical sense, basically means deriving the time-independent single-component disk-like distribution function (DF)
$f\left( \mathbf{x},\mathbf{v} \right)$, of the stars in the phase space mapped with coordinates $\left\{ {{\mathbf{x}},{\mathbf{v}}} \right\}$, or an overlapping set of DFs
$f^{{\rm{tot}}} \left( {{\bf{x}},{\bf{v}}} \right) = \sum {f_j \left( {{\bf{x}},{\bf{v}}} \right)} $ for
each stellar population $j$ that we are able to disentangle.  In terms of
direct observations of the phase space, and within reasonable errors, RAVE
provides the direction (line of sight, l.o.s.) and the velocity along the l.o.s. for each
star. The radial velocity is the observed projection of the true heliocentric observed velocity vector ${\mathbf{v}}_{{\text{hel}}} $
of the star along the line of sight, $v_r  = \left\langle {{\mathbf{v}}_{{\rm{hel}}} ,{\mathbf{\hat r}}_{\rm{hel}} } \right\rangle $ (with $\left\langle { ... , ... } \right\rangle $ denoting the standard inner product).  The direction ${\mathbf{\hat r}}_{\rm{hel}}  \equiv {{{\mathbf{x}} - {\mathbf{x}}_ \odot  } \mathord{\left/
 {\vphantom {{{\mathbf{x}} - {\mathbf{x}}_ \odot  } {\left\| {{\mathbf{x}} - {\mathbf{x}}_ \odot  } \right\|}}} \right.
 \kern-\nulldelimiterspace} {\left\| {{\mathbf{x}} - {\mathbf{x}}_ \odot  } \right\|}}$ is just the heliocentric position vector
divided by the unknown heliocentric distance of each star, $\left\| {{\mathbf{r}}_{\rm{hel}} } \right\| = \left\| {{\mathbf{x}} - {\mathbf{x}}_ \odot  } \right\| \equiv r_{\rm{hel}} $ (with components $\left\{ {\hat x,\hat y,\hat z} \right\}$ in a suitable reference system, see Section \ref{Projectiontechniqueforradialvelocities}).

The analysis of the kinematic properties of the thin and thick disks is of
paramount importance for the comprehension of the origin and formation of
our Galaxy as well as of every disk galaxy. The object of our analysis in
the present paper is mainly the thick disk. A separate paper is dealing with the thin disk analysis \citep[hereafter Paper II,][]{OrangePaper}.

In this paper we investigate the kinematic properties of the Galactic disk
components by applying a methodology based on Singular Value
Decomposition (SVD). SVD permits one to find the solution, in a
least-squares sense, of an inhomogeneous system of N linear equations in the
case of singular matrices (where N is the number of stars we select).  A
number of studies have used this method in the context of proper motion
survey analyses: see, e.g., a fully analytical exercise developed for
HIPPARCOS proper motion and parallax data in \citet{1998MNRAS.298..387D} or
a recent work by \citet{2009AJ....137.4149F} combining data release 7 of
the Sloan Digital Sky Survey \citep{2009ApJS..182..543A} with astrometric
USNO-B data supplemented by photometric
distances.

Finally, several authors devised methods to derive photometric distances for RAVE \citep[e.g.,][]{2008ApJ...685..261K, 2011ApJ...726..103K, 2010A&A...522A..54Z, 2010A&A...511A..90B, 2011A&A...532A.113B}. Here we will adopt the internal release of the catalogue of \cite{2010A&A...522A..54Z} which includes different distance determinations depending on different sets of isochrones for about 260,000 stars \citep[see][for an extended discussion]{2010A&A...522A..54Z}. We will use the Yonsei-Yale distances based on isochrones of \citet[][]{2004ApJS..155..667D}.

The structure of this paper is as follows: in Section
\ref{Preparingthedata} we will prepare the data for the analysis by
cleaning the radial velocities from the effects of Galactic rotation. In
Section \ref{Projectiontechniqueforradialvelocities} we briefly review the
inversion techniques we are going to apply for the analysis of the data. In
Section \ref{results} we present the results and discuss them in Section
\ref{thend}.

\section{Preparing the data}\label{Preparingthedata}

Our procedure requires the following steps.  We need to
\begin{enumerate}
	\item correct the RAVE survey data for large scale effects (this Section \ref{Preparingthedata}),
	\item isolate the thick disk (see Section \ref{Projectiontechniqueforradialvelocities}),
	\item evaluate the thick disk kinematics (see Section \ref{results}).
\end{enumerate}
As an additional result of this procedure, we will be able to estimate
two of the three components of the solar motion relative to the LSR and the velocity lag of the thick disk component.

In this way, we can determine some of the moments of the underlying
multi-component distribution function ${f^{{\text{tot}}} }\left(
\mathbf{x},\mathbf{v} \right)$. While strictly speaking only the infinite series of
the moments is equivalent to the original distribution function, only the
first few moments will be the subject of our exercise. \textit{Thus, our study is
intended to be of an exploratory nature rather than exhaustive, for a set
of problems that are far from being fully solved}.

\subsection{Radial velocity component from the Galaxy differential rotation}
\label{RadialvelocitycomponentfromtheGalaxydifferentialrotation}

For a given position of a star in the Galactic disk, we want to estimate
the influence of the differential rotation of the Galaxy in the radial
velocity component. This is necessary to minimize the influence of the radial component due to the Galactic rotation on the observed radial velocity.

To compute this first step we note that, for a star at a given position ${\mathbf{x}}$
in the Milky Way's disk, the following general vector relation holds:
\begin{equation}\label{EQ01}
{{\mathbf{\bar{v}}}_{\text{LSR}}}\left( {{\mathbf{x}}_{\odot }} \right)+{{\mathbf{v}}_{\odot }}+{{\mathbf{v}}_{\text{hel}}}={{\mathbf{\bar{v}}}_{c}}\left( \mathbf{x} \right)+{{\mathbf{v}}_{p}},
\end{equation}
where ${{\mathbf{\bar{v}}}_{c}}\left( \mathbf{x} \right)$ is the mean
rotational velocity at the given position $\mathbf{x}$ in the Galactic disk reference system centred on the barycentre of the Milky Way (MW) (considered at rest or in rectilinear unperturbed motion).
${{\mathbf{\bar{v}}}_{\text{LSR}}}\left( {{\mathbf{x}}_{\odot }}
\right)$ is the Solar Local Standard of Rest which differs from the Local Standard of Rest (LSR) speed at any other location in the Galaxy.
${{\mathbf{v}}_{\odot }}$ is the Solar peculiar velocity relative
to the Solar LSR and ${{\mathbf{v}}_{p}}$ is the peculiar velocity of the star relative to its own mean Galactic rotational speed.

It is also convenient to introduce here the orthonormal standard system of
reference in the velocity space $\left(O;U,V,W \right)$.  This reference
system is centred on the velocity of the Solar LSR, $O$, with $U$ aligned with
the reference system in the configuration space pointing to the Galactic
centre, $V$ aligned with the rotation of the Galaxy and $W$ pointing to the
north Galactic pole (NGP).  We will call the generic velocity components in
this reference system $\mathbf{v}=\left\{ {{v}_{U}},{{v}_{V}},{{v}_{W}}
\right\}$. From Eqn.  \eqref{EQ01}, if we express the component of the
radial velocity due to Galactic rotation as a function of the star's
longitude and latitude $\left( l,b \right)$, we get
\begin{equation}\label{EQ02}
	\begin{gathered}
v_r^G  = \left( {\left\| {{\mathbf{\bar v}}_c \left( {\mathbf{x}} \right)} \right\|\frac{{R_ \odot  }}{R} - V_{\rm{LSR}} \left( {R_ \odot  } \right)} \right)\cos b\sin l.
\end{gathered}
\end{equation}
For simplicity we assume cylindrical symmetry $\left( O;R,\phi ,z
\right)$ in the configuration space for the Galactic model, ${\mathbf{v}}_{\rm{LSR}} \left( {{\mathbf{x}}_ \odot  } \right) = \left\{ {0,V_{\rm{LSR}} \left( {R_ \odot  } \right),0} \right\}$. ${{\mathbf{v}}_{\odot }}=\left\{ {{v}_{U,\odot }},{{v}_{V,\odot
}},{{v}_{W,\odot }} \right\}$ are the components of the peculiar motion of the Sun
relative to the Solar LSR. $R$ is the Galactocentric radial distance in cylindrical
coordinates for the position of a given star. ${{R}_{\odot }}$ is the
position of the Sun on the plane of symmetry of the MW at an azimuthal
position ${{\phi }_{\odot }}=0$. The height of the Sun relative to the
plane of symmetry of the Galaxy is hereafter neglected, ${{z}_{\odot
}}\cong 0$. The orthonormal rotation matrix ${\mathbf{R}}^T {\mathbf{R}} = {\mathbf{1}}$, i.e.
 $\mathbf{R}\in SO\left( 3 \right)$ with
$\det=+1$ is adopted here to transform a velocity vector (given in terms of
radial velocity ${{v}_{r}}$ and motion along the Galactic coordinates
$\left( {{v}_{l}},{{v}_{b}} \right)$) into a vector in the velocity space reference system vector
$\left( O;U,V,W \right)$. This orthonormal rotation matrix is defined as
\begin{equation}\label{EQ03}
\mathbf{R}=\left(
\begin{matrix} \cos b\cos l & \cos b\sin l & \sin b  \\ -\sin l & \cos l &
0  \\ -\sin b\cos l & -\sin b\sin l & \cos b  \\
\end{matrix} \right).
\end{equation}

\subsection{The influence of distance}
We show now how the influence of the Galactic rotation on the radial
velocity is strongly dependent on the viewing direction, i.e., the
direction $\left( l,b \right)$ for a specific star of the survey, and on
the mean streaming velocity $\left\| {{{\mathbf{\bar{v}}}}_{c}}\left( \mathbf{x}
\right) \right\|$ that will be our free tuning parameter. On the other hand, we can choose suitable directions $(l,b)$ for which the influence of the Galactic rotation on the radial
velocity is weakly dependent on the distance ${{r}_{\text{hel}}}$ of the star along $(l,b)$. This result holds for the range of distances from the Sun in which RAVE dwarf stars are
mainly located, say ${{r}_{\text{hel}}}\le 1.0{\text{kpc}}$ as shown in Appendix C.

In equation \eqref{EQ02} the Galactic radial distance of a template star can be expressed as a function of the heliocentric distance, ${{r}_{\text{hel}}}$.
In particular the the first term in the sum of Eqn.\ \eqref{EQ02} reads
\begin{equation}\label{EQ04}
	\begin{gathered}
  \left\| {{{{\mathbf{\bar v}}}_c}\left( {\mathbf{x}} \right)} \right\|\frac{{{R_ \odot }}}
{R}\cos b\sin l =  \hfill \\
  \frac{{\left\| {{{{\mathbf{\bar v}}}_c}\left( {\mathbf{x}} \right)} \right\|\cos b\sin l}}
{{\sqrt {1 +\frac{{{r_{\text{hel}}}}}
{{{R_ \odot }}} \cos b( \frac{{{r_{\text{hel}}}}}
{{{R_ \odot }}}\cos b - 2\cos l ) } }}. \hfill \\
\end{gathered}
\end{equation}
We will treat the terms $\left\| {{{\mathbf{\bar{v}}}}_{c}}\left( \mathbf{x}
\right) \right\|$ independently later.  For simplicity, we set $\varepsilon
=\frac{{{r}_{\text{hel}}}}{{{R}_{\odot }}}$. For our approximation we only
consider the kinematics in the solar neighbourhood, defined, e.g., as the zone
where $\varepsilon =\frac{{{r}_{\text{hel}}}}{{{R}_{\odot }}}\cong
\frac{1}{8}$. Then it follows from equations \eqref{EQ02} and \eqref{EQ04}
that in this neighbourhood, by expanding the previous
equation in a Maclaurin series in $\varepsilon$, we obtain
$$
\begin{gathered}
  \frac{{\cos b\sin l}}
  {{\sqrt {1 + \varepsilon \cos b(  \varepsilon \cos b - 2\cos l )}    }} \simeq  \hfill \\
  \cos b\sin l +  \hfill \\
  \varepsilon {\cos ^2}b\cos l\sin l + o\left( {{\varepsilon ^2}}      \right) \hfill \\
\end{gathered}
$$
Thus this expansion suggests that   $v_{r}^{\text{G}}$
is primarily dominated by the direction of observation $\left( l,b \right)$ and the
${{\mathbf{\bar{v}}}_{c}}$  value (which differs from star to star and from the
thin disk to the thick disk), but we expect no strong dependence on
the heliocentric distance of the star within our neighbourhood. This is supporting the idea that the error affecting the photometrically determined distances will not be dramatically relevant for the thick disk parameter determinations \citep[see also][]{1990ApJ...357..435C}. At present, we must consider the term ${{\mathbf{\bar{v}}}_{c}}$ as a free
parameter.  We postpone its treatment and its dependence on the
heliocentric distance ${{\mathbf{\bar{v}}}_{c}}\left( \mathbf{x}
\right)={{\mathbf{\bar{v}}}_{c}}\left( {{\mathbf{r}}_{\text{hel}}} \right)$ until
Section \ref{results}, because its behaviour will be treated in the context
of the inversion technique.
Different arguments hold for the thin disk analysis (see Paper II) where the photometric distance errors are the main source of uncertainties for the trend of the velocity ellipsoid in the meridional plane.

\begin{figure}
\resizebox{\hsize}{!}{\includegraphics{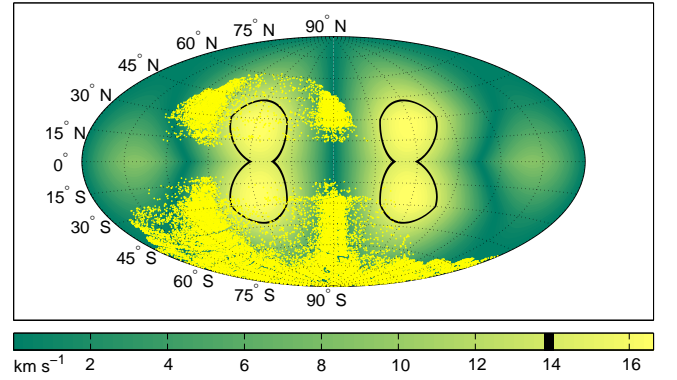}}
\caption{Isocontour plot on an Aitoff projection of the differences in radial velocity due to Galactic rotation $\delta v_r^{\text{G}} $ from Eqn.\ \eqref{EQ06} for test values of
${{\mathbf{\bar{v}}}_{c}}$ calculated using Eqn. \eqref{eqad} and the values in
Table \ref{Tabella01}. The gradients of the velocity dispersion tensor are treated
as in \cite{2006A&A...451..125V}. The assumed values for the motion of the
Sun relative to the LSR are from this paper. The yellow points show the RAVE data selection
from Section
\ref{RadialvelocitycomponentfromtheGalaxydifferentialrotation}. The black
contour lines represent the limit of 14 ${\text{km s}}^{ - 1} $ adopted in our paper.}
\label{SkyDist}
\end{figure}

We can test the validity of the previous approximation without using the expansion into a series, but instead by numerically
solving for the dependence  $v_{r}^{\text{G}}=v_{r}^{\text{G}}\left(
{{r}_{\text{hel}}} \right)$ at different values of $\left( l,b \right)$.
In particular, we are interested in controlling the errors introduced by large scale effects such as differential rotation and photometric distances for selected distances. The dispersion of the thick disk velocity ellipsoid is generally assumed to be constant and thus should not introduce any adverse large scale effects on the small volume considered. Nevertheless, we will check this assumption ``a posteriori'' in Section \ref{futuroeoltre}.
If we limit ourselves to consider the RAVE stars within
a sphere of radius $1.0$ kpc centred at the Sun's location, from Eqn.\eqref{EQ04} we can calculate the value of the radial velocity due to the Galactic rotation $v_r^{\rm{G}} $
along the line of sight $r_{{\rm{hel}}}  \in \left[ {0,1} \right]$ kpc for each direction in the sky specified by $(l,b)$. To evaluate this equation, we test our initial guess values for Eqn. \eqref{EQ04} with the help of a Galactic population model by \cite{2006A&A...451..125V} where the MW Galactic potential and velocity dispersion profiles are consistently determined (see also our Eqn. \eqref{eqad} and Section \ref{aposteriori}).

This value of the Galactic component of the radial velocity ranges from a maximum, $\max \left( {v_r^{\rm{G}} } \right)$, and a minimum value, $\min \left( {v_r^{\rm{G}} } \right)$, whose difference we will call
\begin{equation}\label{EQ06}
\delta v_r^{\rm{G}}  \equiv \left| {\max v_r^{\rm{G}}  - \min v_r^{\rm{G}} } \right|_{r_{{\rm{hel}}}  \in \left[ {0,1} \right].}
\end{equation}
This difference $\delta v_r^{\rm{G}}$ can be evaluated at different values of $\left( l,b \right)$ and for an optimized value of ${{\mathbf{\bar{v}}}_{c}}$. In this way we obtain the background in Figure \ref{SkyDist} that shows the Aitoff sky projection contour plot obtained from Eqn. \eqref{EQ06}.
As we can see from this contour plot, the highest $\delta v_r^{\rm{G}}$ (yellowish zones) occur at low latitude and the smallest in the green zones. For $b=0$, neglecting the motion of the Sun relative to the Solar LSR, Eqn.\ \eqref{EQ06} is
just the terminal velocity relation \citep[e.g.,][]{1998gaas.book.....B}.
The highest induced error is $|\delta v_{r}^{\text{G}}|\cong 16{\text{km s}}^{ - 1} $ at a distance of $1.0{\text{kpc}}$. This may seem
quite large if compared with the mean radial velocity error of RAVE data but this error occurs at low
latitude regions that are not of interest for RAVE because these are not covered
by the survey (see Figure \ref{SkyDist}). Also the dependence of ${{\mathbf{\bar{v}}}_{c}}$ on the range of interest of
${{\mathbf{r}}_{\text{hel}}}$ is again more relevant at low latitude.

\subsection{RAVE data selection criteria}\label{dataselection}
It becomes evident from the previous considerations that when we introduce an indetermination in the distance $d_i$ of the $i^{th}$-star due to the error in the photometrically determined distance $\Delta d_i$, we are also introducing an error in the correction for the radial velocity contribution due to the Galactic rotation. If we proceed with a Monte Carlo approach by randomly generating a distance $d_i$ within $\Delta d_i$ for the $i^{th}$-star, the correction $v_{r,i}^{\rm{G}} $ for this star will assume only values between the two extremes $
{\mathop {\max }\limits_{r_{{\rm{hel}}}  \in \left[ {0,1} \right]} \left( {v_r^{\rm{G}} } \right)}$ and ${\mathop {\min }\limits_{r_{{\rm{hel}}}  \in \left[ {0,1} \right]} \left( {v_r^{\rm{G}} } \right)}$ in Eqn. \eqref{EQ06}.
If this correction is smaller than the intrinsic error on the radial velocity as derived by RAVE, $\Delta v_{r}^{\text{RAVE}}\left( {{l}_{i}},{{b}_{i}}
\right)$, i.e., $\Delta v_{r}^{\text{RAVE}}\left( {{l}_{i}},{{b}_{i}} \right)> \delta v_{r}^{\text{G}}\left( {{l}_{i}},{{b}_{i}} \right)$, we can either apply this correction or we can safely neglect it. This is because its contribution to $v_{r,i}^{\rm{G}} $ is too small to bias our consideration, i.e. its effect cannot be accounted for within the error $\Delta v_{r}^{\text{RAVE}}\left( {{l}_{i}},{{b}_{i}}
\right)$ of the observed stellar radial velocity. Equivalently, by shifting the $i^{th}$-star by its distance uncertainty, it does not affect the radial velocity contribution $v_r^{\text{G}}$ due to the Galactic rotation because it is negligible compared to the intrinsic error $\Delta v_{r}^{\text{RAVE}}\left( {{l}_{i}},{{b}_{i}} \right)$. We point out that the observational errors, $\Delta v_r$, for which we
use capital delta ($\Delta$), are of a different nature than the errors $\delta v_r^{\text{G}}$ described by Eqn.\ \eqref{EQ06}, which have different systematic trends and an origin that can be accounted for with a complete Galactic dynamics model \citep[see][their Appendix]{2006A&A...451..125V}, and our Eqn. \eqref{eqad}).

In order to adopt an even more restrictive condition, we will use all those stars from the RAVE catalogue that have an
intrinsic error $\Delta v_{r}^{\text{RAVE}}\left( {{l}_{i}},{{b}_{i}}
\right)$ greater than twice the error introduced by
neglecting the stars' distance dependence on the velocity due to the
Galactic rotation. In other words, for every star $i$, we require:
\begin{equation}\label{EQ07}
	\Delta v_{r}^{\text{RAVE}}\left( {{l}_{i}},{{b}_{i}} \right)>2.0 \delta v_{r}^{\text{G}}\left( {{l}_{i}},{{b}_{i}} \right).
\end{equation}
A test of the validity of all these arguments to recover the correct results has been performed on a mock catalogue as explained in Section \ref{Stabilityofresults}.

 From the sample of stars that survive the selection defined by Eqns. \eqref{EQ06} and \eqref{EQ07} shown in Figure \ref{SkyDist}, we proceed further with a few extra selection criteria. We consider stars with signal-to-noise ratio $S/N > 20$. We require the near-infrared colours of the RAVE stars taken from 2MASS observations to be in the range of $J - K \in \left[ {0.2,1.1} \right]$ in order to clean the data set from extremely young stars. Furthermore, we require $v_r^{\text{RAVE}}  < 300 {\mbox{
}\text{km s}}^{{-1}}  $ to avoid stars dynamically not representative of the thick disk within 2 kpc from the Sun's position, $\Delta v_r^{\text{RAVE}}  < 14 {\mbox{
}\text{km s}}^{{-1}} $ to reduce the propagation error while retaining a considerable amount of stars (see the velocity contour highlighted in black in Fig. \ref{SkyDist}), $\left| b \right| > 10^\circ $ because we are not interested in thin disk stars, and $\log _{10} g > 3.5$ to avoid the influence of giant stars which may sample a much more distant part of the Galaxy. Together with Eqn. \eqref{EQ07} the last criterion represents the most severe cut on the total number of stars that we can use. Where possible a further cut
$\left[ {\text{Fe}/\text{H}} \right] > -1$ dex
was applied adopting the metallicity determination described by \cite{2008AJ....136..421Z}.

Starting with an initial number of roughly 260,000 stars in the
data release of \cite{2010A&A...522A..54Z}, the final, remaining number of stars for which we can perform
the analysis is $N_{\text{tot}}\simeq 38,805$ which is orders of magnitude larger than the samples currently available in the literature (see Fig. \ref{Figure05} for a plot of the error distribution).

\section{Inversion techniques for radial velocities}\label{Projectiontechniqueforradialvelocities}

In the following Section \ref{sguardo} we will present the technique to determine the principal moments of a DF for a mixture of thin and thick disk stars from a radial velocity survey. With a few extra assumptions we will be able to disentangle the two components and focus our attention on the thick disk as will be shown in Section \ref{spaccotutto}.

\subsection{The mixture of the thin and thick disk distribution of moments}\label{sguardo}
Once we have selected the sample of RAVE stars that satisfy the conditions laid out in Section \ref{dataselection} (see yellow dots in Fig. \ref{SkyDist}), we can deduce the first moments of the underlying mixture distribution function with simple algebra based on a very
popular technique for dealing with sets of linear equations
that can be written in the matrix form ${\mathbf{Ax}} = {\mathbf{b}}$, for the unknown vector $\mathbf{x}$. In the following we recall the basics applied to our specific case
of a radial velocity survey. For more details we refer to Appendix A.

Based on Eqn.\ \eqref{EQ01} we can express the components of the radial
velocity vector, say ${{\mathbf{v}}_{\parallel }}$, by
\begin{equation}\label{EQ09}
{{\mathbf{v}}_{\parallel }}=\left( \begin{matrix}
   {{v}_{r}}\cos b\cos l  \\
   {{v}_{r}}\cos b\sin l  \\
   {{v}_{r}}\sin b  \\
\end{matrix} \right),
\end{equation}
where ${{\mathbf{v}}_{\parallel }}=\left\langle \mathbf{v},\mathbf{\hat{r}}
\right\rangle \mathbf{\hat{r}}\equiv \mathbf{p}\mathbf{v}$ with $\mathbf{p}$ being the
idempotent matrix of the projection operator along the line of sight. Proceeding component by component we obtain:
\begin{equation}\label{EQ10}
{\mathbf{v}}_\parallel   = {\mathbf{p}\mathbf{v_{\text{hel}}}} = {\mathbf{\hat r}}^{ \otimes 2} {\mathbf{v_{\text{hel}}}} = \left( {\begin{array}{*{20}c}
   {\hat x^2 } & {\hat x\hat y} & {\hat x\hat z}  \\
   {\hat x\hat y} & {\hat y^2 } & {\hat y\hat z}  \\
   {\hat x\hat z} & {\hat y\hat z} & {\hat z^2 }  \\

 \end{array} } \right).\left( {\begin{array}{*{20}c}
   {v_U }  \\
   {v_V }  \\
   {v_W }  \\

 \end{array} } \right),
\end{equation}
where $\left( {...} \right) \cdot \left( {...} \right)$ is the ordinary matrix product, ${\mathbf{a}}^{ \otimes n} $ is the standard tensor power of the generic
vector ${\mathbf{a}}$  (and $n \in \mathbb{N}$ any non-negative integer) and
$\mathbf{\hat{r}}$ is again the unitary vector of a star whose components are
$\left\{ \hat{x},\hat{y},\hat{z} \right\}$ in the configuration space
collinear with the velocity space $\left( O;U,V,W
\right)$ and centred on $O$. Clearly the matrix $\mathbf{p}$ is singular (its $\det \mathbf{p}=1-{{\left\|
{\mathbf{\hat{r}}} \right\|}^{2}}=1-1=0$) thus not permitting us to determine its
inverse.

Nevertheless, we can proceed by taking into consideration the whole subsample of the selected RAVE data in order to statistically invert the overdetermined system ${\mathbf{Ax}} = {\mathbf{b}}$ that we obtain by defining ${\mathbf{A}}$ as the block diagonal matrix of all the projection operators ${\mathbf{p}}$ once the off-diagonal blocks are small (see Appendix A for further details). We call this (3Nx3) block matrix of all the projection operators simply ${\mathbf{P}}$. ${\mathbf{b}}$ will be defined from the vectors of the observed radial velocities for the selected sample, which we call for simplicity again ${\mathbf{v}}_\parallel  $. We obtain in this way the system ${\mathbf{P\bar v}}_{{\rm{hel}}}  = {\mathbf{v}}_\parallel  $ that we multiply as usual by the transposed matrix ${\mathbf{P}}^T $ in order to obtain the square matrix product ${\mathbf{P}}^T {\mathbf{P}}$ that can be inverted in order to give
$ {\mathbf{\bar v}}_{{\rm{hel}}}  = \left( {{\mathbf{P}}^T {\mathbf{P}}} \right)^{ - 1} {\mathbf{P}}^T {\mathbf{v}}_\parallel   \equiv {\mathbf{P}}^ +  {\mathbf{v}}_\parallel $, where ${\mathbf{P}}^ +$ is the pseudo inverse of ${\mathbf{P}}$ \citep{1956PCPS...52...17P,1955PCPS...51..406P}.
Probably the most widely used orthogonal decomposition suitable for the solution of our system is the Singular Value Decomposition due to its numerical stability \citep[e.g.,][]{1986nras.book.....P}, which we will also adopt here (e.g., by writing ${\mathbf{P}}^ + \equiv \frac{{{\mathbf{WU}}^T }}
{{diag{\mathbf{Q}}}}$ with $\mathbf{W}$ and $\mathbf{U}$ as orthogonal matrices and $\mathbf{Q}$ as a diagonal matrix (see e.g., \citet{1986nras.book.....P} for an extended discussion and implementation techniques). Finally, in order to check the consistency of this approximation we will keep track of the sky distribution by weighting the obtained matrix ${\bf{P}}$ with the isotropic case laid out in Appendix A (see Sect. \ref{Stabilityofresults} and Eqn. A.3).

\begin{figure}
\resizebox{\hsize}{!}{\includegraphics{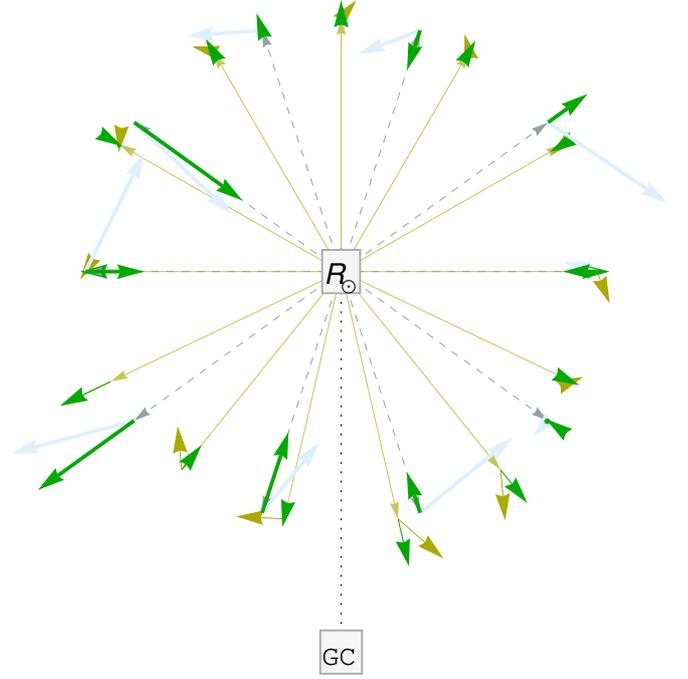}}
\caption{Schematic representation of the hypothesis of an anisotropic and isothermal distribution of an unbiased sample of thick disk velocities superimposed on a thin disk sample. The observer is located at the Sun's position $R_{\odot}$ at the centre. The Galactic centre (GC) is at the bottom of the figure. The directions on the celestial sphere to each star $s_i$ are indicated with dashed arrows for thick disk stars and solid arrows for the thin disk stars. The green thick arrows represent the radial velocities of the thick disk, whose corresponding unknown true velocity vector is in light blue. The thin disk radial velocities are represented by the thin green arrows, and the corresponding unknown olive-green arrows represent the true velocity vectors. Note that the olive-green arrows are generally longer in the GC direction and shorter in the anti-Galactic centre direction as expected from the velocity dispersion trend.}
\label{Figure000}
\end{figure}

There is an implicit hypothesis that is underlying our approach. When we are considering the system of
equations ${\mathbf{P} \mathbf{\bar v}}_{{\rm{hel}}} $, we need to implicitly assume that either the group of stars we are sampling is so extremely local that it is not (or only weakly) influenced by the spatial trend of the velocity dispersion tensor, or that the population we are sampling has an isothermal dispersion tensor within the distances sampled. In the latter case the sample does not need to be local. For the thick disk stars we can exploit the latter, while the first has been already exploited in the work of \citet{1998MNRAS.298..387D}. A bi-dimensional schematic representation of this assumption can be seen in Fig. \ref{Figure000}: the light blue arrows represent the unknown velocity vectors of which we observe the radial velocities (green arrows) along a given direction (grey arrows). The velocity dispersion described by the light blue arrows is found (see below)  not to depend on the position inside or outside the solar radius $R_ \odot  $ throughout the volume covered by our sample. For comparison see Figure 2 of Paper II where the trend of the thin disk velocity dispersion tensor along the meridional plane is considered.

This complex procedure provides our first moment of the
composite distribution function, hereafter simply $\mathbf{\bar{v}}$.
With this approach the solar motion relative to a selected sample of stars is
simply the mean motion of the stars relative to the Sun with a change in
sign for the U and W directions, say ${{(v_U,v_W)}_{\odot
}}=-(\overline{v_U},\overline{v_W})$. In the V direction the lag of the
component motion has to be taken into account with an extra $v_{\text{lag}}$ whose
contribution overlaps with $v_{V,\odot}$ to form the mean value
$\overline{v_V}$.

In the same way we can get the second, third, and fourth central moment, by defining
$\mathbf{{v}'}\equiv \mathbf{v}-\mathbf{\bar{v}}$ and averaging, e.g., for the moment of
order 2:
\begin{equation}\label{EQ13a}
\overline {{\mathbf{v'}}_\parallel  ^{ \otimes 2} }  = \overline {\left( {{\mathbf{P}\mathbf{v'}}} \right)^{ \otimes 2} }  = \overline {{\mathbf{P}}^{ \otimes 2} {\mathbf{v'}}^{ \otimes 2} }.
\end{equation}
Hereafter we exploit  the symmetries of the matrices:
\begin{equation}\label{EQ13}
	\bm{\sigma }_{\parallel }^{2}=\frac{1}{2}\overline{\left( \mathbf{P}\otimes \mathbf{P} \right)+{{\left( \mathbf{P}\otimes \mathbf{P} \right)}^{T}}}\cdot {{\bm{\sigma }}^{2}}\equiv {\mathbf{\bar M}}_{\left[ 2 \right]}\cdot {{\bm{\sigma }}^{2}}
\end{equation}
where $\bm{\sigma }_{\parallel }^{2}$ is the dispersion tensor of the
radial velocity and ${{\bm{\sigma }}^{2}}$ is the ordinary pressure tensor.
In the same way, we define the projected third moment along the l.o.s. as
\begin{equation}\label{EQ13III}
{\mathbf{S}}_\parallel   \equiv \overline {{\mathbf{v'}}_\parallel  ^{ \otimes 3} }  = \overline {{\mathbf{P}}^{ \otimes 3} {\mathbf{v'}}^{ \otimes 3} }  \equiv {\mathbf{\bar M}}_{\left[ 3 \right]}  \cdot {\mathbf{S}},
\end{equation}
and the projected fourth moment as
\begin{equation}\label{EQ13IV}
{\mathbf{T}}_\parallel   = \overline {{\mathbf{v'}}_\parallel  ^{ \otimes 4} }  = \overline {{\mathbf{P}}^{ \otimes 4} {\mathbf{v'}}^{ \otimes 4} }  \equiv {\mathbf{\bar M}}_{\left[ 4 \right]}  \cdot {\mathbf{T}},
\end{equation}
where the suitable symmetries have been exploited.
The interested reader can find the element by element characterization of the matrices ${\mathbf{\bar M}}_{\left[ i \right]} $ in Appendix A.

The errors in these quantities are given by standard statistical tools
coming from linear regression, taking into account that each star $i$
contributes to ${{\mathbf{v}}_{\parallel }}$ with its own error $\Delta
{{\mathbf{v}}_{\parallel i}}$ (referred to as $\Delta v_{r}^{\text{RAVE}}\left(
{{l}_{i}},{{b}_{i}} \right)$ in Eqn.\ \eqref{EQ07}).  Hence, for instance,
the error on the mean will be $\Delta \mathbf{\bar{v}}^{2}={{\left(
\overline{{{\mathbf{P}}^{+}}} \right)}^{2}}\overline{\Delta {{\mathbf{v}}^{2}}},$
on the dispersion tensor it will be
	$\Delta \bm{\sigma }^2=2{{\left( \overline{{{\left( \mathbf{P}\otimes \mathbf{P} \right)}^{+}}} \right)}^{2}}\overline{{{\left( {{{\mathbf{{v}'}}}_{\parallel }}\otimes \Delta {{{\mathbf{{v}'}}}_{\parallel }} \right)}^{2}}}$,
and so on (exactly as in  \citet{2009AJ....137.4149F}).

\subsection{The scattering processes and the thick disk component disentanglement}
\label{spaccotutto}

So far, we have a procedure that, through the Eqns. \eqref{EQ13a}, \eqref{EQ13}, \eqref{EQ13III} and \eqref{EQ13IV}, provides us with the first four true moments (i.e., $\left\{ {{\mathbf{\bar v}},{\bm{\sigma }},{\mathbf{S}},{\mathbf{T}}} \right\}
$) of the
distribution function of the thin and thick disk mixture from the projected moments along the l.o.s. (i.e., $\left\{ {{\mathbf{\bar v}}_\parallel  ,{\bm{\sigma }}_\parallel  ,{\mathbf{S}}_\parallel  ,{\mathbf{T}}_\parallel  } \right\}$) that we obtain directly from the observations.

To proceed further
with the thick disk analysis, we want to disentangle the two distribution
functions (DFs) and proceed by determining the thick disk velocity dispersion tensor alone.

We assume that the two single-particle DFs for the thin and thick disk,
$f_1 \left( {{\mathbf{x}},{\mathbf{v}}} \right) \equiv f_{\text{thin}} $
and $f_2 \left( {{\mathbf{x}},{\mathbf{v}}} \right) \equiv f_{\text{thick}}$,
can be added linearly to yield the overall distribution function
$f^{{\text{tot}}}  \left( {{\mathbf{x,v}}} \right) = f_1  + f_2 $.  For the mixture distribution function $f^{{\text{tot}}} $ in the previous
Section we computed the first four moments. The first and the second moments
correspond to the mean of the sample and to the dispersion tensor, respectively,
\begin{equation}\label{EQs}
	{\bm{\sigma }^2} = \frac{1}
{N}\int_{}^{} {{\mathbf{v'}}^{ \otimes 2} f^{{\text{tot}}}  \left( {{\mathbf{x,v}}} \right)d{\mathbf{v}}} ,
\end{equation}
where the integral extends over the whole velocity space and $N$ is the number of stars. The third
and fourth moments are:
\begin{equation}\label{EQ3}
{\mathbf{S}} = \frac{1}
{N}\int_{}^{} {{\mathbf{v'}}^{ \otimes 3} f^{{\text{tot}}}  \left( {{\mathbf{x,v}}} \right)d{\mathbf{v}}} ,
\end{equation} and
\begin{equation}\label{EQ4}
	{\mathbf{T}} = \frac{1}
{N}\int_{}^{} {{\mathbf{v'}}^{ \otimes 4} f^{{\text{tot}}}  \left( {{\mathbf{x,v}}} \right)d{\mathbf{v}},}
\end{equation}
with the same definition interval for the integrals. We see that $N =
N\left( {{\mathbf{x}},t} \right)$, as well as ${\bm{\sigma }^2} = {\bm{\sigma
}^2}\left( {{\mathbf{x}},t} \right)$, ${\mathbf{S}} = {\mathbf{S}}\left( {{\mathbf{x}},t}
\right)$ and ${\mathbf{T}} = {\mathbf{T}}\left( {{\mathbf{x}},t} \right)$ because, in
general, $f^{{\text{tot}}} = f^{{\text{tot}}}  \left( {{\mathbf{x}},{\mathbf{v}},t}
\right)$. The form of this distribution function is generally unknown but
its sub-components, $f_{\text{thin}} $ or $f_{\text{thick}} $, are often
assumed to be quadratic in the peculiar velocity components,
i.e., of a generalized Schwarzschild type.

In the time evolution of the total DF $f^{{\text{tot}}} $, we can argue
that a very different role is played by $f_{\text{thin}} $ or
$f_{\text{thick}} $.
A few processes of scattering are generally assumed to influence
the time evolution of $f_{\text{thin}}$: mostly the scattering due to transient spiral
arms and the encounters with giant molecular clouds \citep[e.g.,][]{1987gady.book.....B}. Both processes lead to an increase of the
velocity dispersion with increasing distance from the plane and provide insight into the
origin of the age-velocity dispersion relation. Moreover, the Schwarzschild
characterization  of $f_{\text{thin}}$, on large scales in the velocity space, is
not as good a description on smaller scales \citep[e.g.,][]{2007MNRAS.380.1348S}. Sub-structures are mainly
found for young stars (but not only for these). Such stars in a moving
group are born at the same place and time, and then disperse into a stream
that may intersect the solar neighbourhood. In this scenario, the stars that
are moving in the same group should share the same age, metallicity, and
azimuthal velocity \citep[e.g.,][]{1998AJ....115.2384D,
1998A&A...340..384C}. A different explanation for the sub-structures in the velocity DF is given
by \citet{2004MNRAS.350..627D}. These authors suggest that sub-structures
arise naturally from the same spiral gravitational fluctuations that excite
the growth of the velocity dispersion. In this picture, sub-structures are
caused by homogeneous star formation in an irregular potential, as opposed
to inhomogeneous star formation in a regular potential. Once the thin disk component is disentangled from the thick disk component, its analysis will proceed with the full exploitation of the proper motions (Paper II).

The situation is less clear for the thick disk where the role of the scattering processes is still undetermined. Its formation may be related to the influence of massive satellites that can either heat pre-existing disks (preserving a vertical metallicity gradient if present, \citep{2011A&A...525A..90K} or get accreted \citep[e.g.,][]{1986ApJ...309..472Q, 1993ApJ...403...74Q}. Other formation scenarios include gas-rich mergers \citep[e.g.,][]{2004ApJ...612..894B}, or
internal radial migration processes  \citep[e.g.,][]{2009MNRAS.396..203S} or formation induced by satellites \citep[e.g.,][]{2011A&A...525L...3D}. We will start working with the hypothesis that the thick disk DF can be approximated as $f_{\text{thick}}  \propto e^{ - \frac{1}
{2}\left( {{\mathbf{v'}}^T {\mathbf{C}}_{\text{thick}} {\mathbf{v'}}} \right)} $ where
the covariance matrix ${\mathbf{C}}_{\text{thick}}^{ - 1} $ is ${\bm{\sigma
}^2}_{\text{thick}} $ and the quadratic equation  ${\mathbf{v'}}^T
{\mathbf{C}}_{\text{thick}} {\mathbf{v'}} = 1$ defines a velocity ellipsoid that
characterizes the distribution function of the thick disk fully and that we
want to determine. We will check ``a posteriori'' (see Section \ref{futuroeoltre}) if this approximation is consistent with our results for the group of stars selected as in Section \ref{dataselection}. The thick disk is more diffuse in the solar neighbourhood
than the thin disk and is characterized by a larger scale height
\citep[e.g.,][]{1998gaas.book.....B}.  We assume that the dependence of the velocity ellipsoid in the sample analysed with RAVE can be adequately
considered to be independent from the Galactic plane distance, i.e., an
isothermal but anisotropic DF, as commonly assumed in most of the thick disk
velocity ellipsoid determinations available in the literature \citep[e.g.,][]{1996AJ....112.2110L, 2005A&A...442..929A, 2000AJ....119.2843C,
2007A&A...475..519H}. We will return to the validation of this hypothesis in Section \ref{Discussionandconclusion}.

Under these assumptions, the decoupling between the thin and thick disk
component is straightforward and analytically developed in, e.g.,
\citet{1992AJ....103.1608C} (see also \citet{2007MNRAS.380..848C}). We are
going to apply this inversion technique in its version based on the cumulants
(\citet{2004A&A...427..131C}) because this is numerically more stable and
hence suitable to minimize the numerical error propagation.

We outline the methodology applied here because it differs slightly from
the original work of \citet{2004A&A...427..131C}. The interested
reader is referred to Appendix B, where the technique contained in
\citet{2004A&A...427..131C} is explicitly derived and adapted for our specific case.  The derivation of the cumulants from the
Eqns.\  \eqref{EQs},  \eqref{EQ3} and \eqref{EQ4} is straightforward:
\begin{equation}\label{EQCum}
\begin{gathered}
  {\bm{\kappa }}^{\left[ {\bm{\sigma }} \right]}  = \frac{N}
{{N - 1}}{\bm{\sigma }}   \hfill \\
  {\bm{\kappa }}^{\left[ {\mathbf{S}} \right]}  = \frac{{N^2 }}
{{\left( {N - 1} \right)\left( {N - 2} \right)}}{\mathbf{S}} \hfill \\
  {\bm{\kappa }}^{\left[ {\mathbf{T}} \right]}  = \frac{{N^2 \left( {N + 1} \right)}}
{{\left( {N - 1} \right)\left( {N - 2} \right)\left( {N - 3} \right)}}\left( {{\mathbf{T}} - \frac{{N - 1}}
{{N + 1}}{\bm{\sigma }}^{\otimes 2}} \right),   \hfill \\
\end{gathered}
\end{equation}
hereafter simply indicated as $\kappa_{ij}$, $\kappa_{ijk}$,
$\kappa_{ijkl}$, respectively.  If we let $N_{\text{thin}} $  and
$N_{\text{thick}} $  be the unknown numbers of stars for the thin and thick
disk components, we can define the parameter $q$ as follows:
\begin{equation}\label{EQq}
q \equiv \sqrt {\frac{{n_{\text{thick}} }}
{{n_{\text{thin}} }}}  - \sqrt {\frac{{n_{\text{thin}} }}
{{n_{\text{thick}} }}},
\end{equation}
where  $n_{\text{thin}}  = {\raise0.7ex\hbox{${N_{\text{thin}} }$}
\!\mathord{\left/ {\vphantom {{N_{\text{thin}} }
N}}\right.\kern-\nulldelimiterspace} \!\lower0.7ex\hbox{$N$}}$ and
$n_{\text{thick}}  = {\raise0.7ex\hbox{${N_{\text{thick}} }$}
\!\mathord{\left/ {\vphantom {{N_{\text{thick}} }
N}}\right.\kern-\nulldelimiterspace} \!\lower0.7ex\hbox{$N$}}$ are the
unknown fractions of stars in the thin and thick disk. This parameter
varies from 0 when $50\% $  of the stars belong to each component to
$\infty $ when $100\% $ of the stars belong to the thick disk.
\textit{Our goal is to maximize this parameter }$q$  in the subsample of
stars selected from the RAVE catalogue as in Section \ref{dataselection}.

\citet{2004A&A...427..131C} show the great advantage of exploiting the
symmetries of the normal distribution functions by working with a
transformed peculiar vector. We follow this convention by defining a
normalized vector as
\begin{equation}
{\mathbf{\hat d}} \equiv \sqrt {n_{\text{thin}} n_{\text{thick}} } \frac{{\mathbf{w}}}
{{w_2 }},
\label{EQdvect}
\end{equation}
where ${\mathbf{w}} = \left\{ {w_1 ,w_2 ,w_3 } \right\}$  is the velocity vector of the difference between the thin and thick disk centroids; ${\mathbf{w}} =
{\mathbf{\bar v}}_{\text{thin}}  - {\mathbf{\bar v}}_{\text{thick}} $. The reason
for the normalization to the highest velocity difference component of
$\mathbf{w}$ is to reduce the error propagation in the SVD solution using the
best-determined differential velocity, i.e., $w_2$.  We define the
transformed peculiar velocity ${\mathbf{V}}$ that keeps the differential
velocity in the azimuthal direction invariant:
\[
\begin{gathered}
  {\mathbf{V}} = {\mathbf{H}} \cdot {\mathbf{v}} \hfill \\
  {\mathbf{H}} = \left( {\begin{array}{*{20}c}
   0 & { - \hat d_3 } & 1  \\
   0 & 1 & 0  \\
   { - 1} & {\hat d_1 } & 0  \\
 \end{array} } \right) \hfill \\
\end{gathered}
\]
This is an isomorphic transformation of the peculiar velocity vector
${\mathbf{v}}$, with $\det \left( {\mathbf{H}} \right) = 1$ due to the previously
assumed normalization of Eqn.\ \eqref{EQdvect}.  The new cumulant equation
for the transformed DF function of ${\mathbf{V}}$ can be derived as a function
of the old cumulants of Eqn.\ \eqref{EQCum} (see Appendix B). Once the new
${\mathbf{V}}$-mixture cumulants have been computed, say $ {\mathbf{{\rm
K}}}^{[3]}$ and ${\mathbf{{\rm K}}}^{[4]} $ for the third and fourth cumulants respectively, we determine the parameter $q$ of Eqn. \eqref{EQq} by making again use of the SVD technique to determine the solution of the linear system of over-determined equations as already done before for the system of equations in Section \ref{sguardo}. We present the description of the system in Appendix B to which we refer the interested reader.

Finally we extrapolate the values of $n_{\text{thick}} $,
${\bm{\sigma }}_{\text{thick}} $, and $w_2 $, i.e. the characterizing thick disk kinematics parameters we were searching for, with a simple numerical iterative procedure that maximizes the thick disk component alone \citep[e.g., Simulated Annealing Methods,][]{1986nras.book.....P}. Within the preselected set of data satisfying the selection criteria of Section \ref{dataselection}, we consider a Monte Carlo generation of the stellar  parameters within their observational errors and we isolate the thick disk component alone with the methodology described above. The kinematic characterization of the thick disk is achieved with the same purely geometrical approach developed for the mixture but now on the thick disk selected sample alone. Once we apply this Monte Carlo approach to generate several catalogues, and then apply the procedure  to each of them, we summarize our results after a simple statistical analysis.

\section{Results}\label{results}
Here we present the results for the determination of the motion of the Sun relative to the LSR and the velocity dispersion tensor for the thick disk.

\subsection{Solar motion relative to the LSR and thick disk rotation velocity}
\label{SmLSR}

Computing the first moment of the velocity part of
the distribution function, namely the mean, we are left with a free
parameter $\left\| {{{\mathbf{\bar{v}}}}_{c}} \right\|$. We assume that the
dependence of  $\left\| {{{\mathbf{\bar{v}}}}_{c}} \right\|$ on the model is
rather weak for the two components, ${{v}_{U,\odot }}$ and ${{v}_{W,\odot
}}$, which we hence derive as
\begin{equation}\label{solmot}
\begin{array}{l}
 v_{U, \odot }  = \left( {9.87 \pm 0.37} \right){\text{km s}}^{ - 1}  \\
 v_{W, \odot }  = \left( {8.01 \pm 0.29} \right){\text{km s}}^{ - 1}.  \\
 \end{array}
\end{equation}
The determination of the last value of ${{\mathbf{\bar{v}}}_{\odot }}$:
${{v}_{V,\odot }}$ is more complicated. This is mostly for the following
two reasons:
\begin{enumerate}
	\item The overlap of the rotational delay of the population ${{v}_{\text{lag}}}$ and the contribution of the peculiar motion of the Sun in the same direction ${{v}_{V,\odot }}$, both
acting together on the mean streaming velocity $\left\|
{{{\bar{\mathbf{v}}}}_{c}} \right\|$, prevents us from obtaining the relative
velocity between the LSR and the standard of rest centred on the motion of
the Sun from our first distribution moment, $\mathbf{\bar{v}}$.  Thus ${{v}_{V,\odot }}$ remains undetermined.
	\item A full determination of ${{\mathbf{\bar{v}}}_{\odot }}$ could in
principle be achieved with this methodology by including all stars down to a low latitude, say,
$b<15$ deg.  However, the survey data are missing this low-latitude sky
coverage.  Moreover, there is no mapping in the Galactic rotation
direction (see Figure \ref{SkyDist}).
\end{enumerate}
Nevertheless, while we cannot carry out a clean determination of the first
moment of the velocity section of the DF, by assuming the literature value
${{\hat{v}}_{V,\odot }}=13.5 \pm 0.3{\mbox{ }\text{km s}}^{-1}$ from
\cite{2009NewA...14..615F}, we gain a value for the velocity lag of our
thick disk sample of stars,
\begin{equation}\label{vlag}
	{{v}_{\text{thick,lag}}}\cong 49 \pm 6{\mbox{ }\text{km s}}^{-1}.
\end{equation}

As explained at the end of the previous section, the procedure works iteratively in order to select a single population of thick disk stars from which to deduce the kinematic parameters. This is done because we cannot determine the solar motion relative to the Solar LSR from a non-local mixture because the thin disk velocity dispersion trend on the meridional plane could bias the results. Thus we need to disentangle one single component from which to derive $\left( {v_{U, \odot } ,v_{W, \odot } } \right)$.

\subsection{Thick disk velocity ellipsoid}

The values determined for the velocity dispersion tensor, ${{\sigma
}_{RR}}=59.2 \pm 4.4\ {\mbox{ }\text{km s}}^{-1}$, ${{\sigma }_{\phi \phi
}}=47.3 \pm 7.5\ {\mbox{ }\text{km s}}^{-1}$ and ${{\sigma }_{zz}}=35.9 \pm
4.1\ {\mbox{ }\text{km s}}^{-1}$ (see Table \ref{Tabella01}), are in good agreement with what was
already presented in the literature for the thick disk component \citep[e.g.,][]{2005A&A...442..929A, 2000AJ....119.2843C, 2007A&A...475..519H}. The
presence of the thin disk is expected to bias these values only slightly in
the sense that the random selection of the stars used in order to maximize
the parameter $q$ in Eqn.\eqref{EQq} is proven to be able to separate the two components as
shown in \cite{2004A&A...427..131C} and as confirmed by the simulated mock
catalogues (see Section \ref{Stabilityofresults}).

As can be seen, the vertical
tilt of the velocity ellipsoid is deduced from the mixed term component
${{\sigma }_{Rz}}=10.1 \pm 3.3\ {\mbox{ }\text{km s}}^{-1}$, leading to a
tilt angle of about $\phi =\frac{1}{2}\arctan \left( \frac{2\sigma
_{Rz}^{2}}{\sigma _{RR}^{2}-\sigma _{zz}^{2}} \right)\cong 3.07\pm 1.1\deg
$. This is similar to the result of \citet{2008MNRAS.391..793S}, who
applied different selection criteria to the same data, but deviates from
the recent determination by \citet{2009AJ....137.4149F}. Regarding the
comparison with these studies it is worth noting that we used Eqn.\ \eqref{EQ07}, while in
\citet{2009AJ....137.4149F} the tilt depends on the mixture of the two components (thin and thick disk)  and it was considered
without any technique capable of disentangling the relative
statistical influence of the two, nor accounting for the radial dependence on $R$ in $\sigma_{ij}=\sigma_{ij}(R,z)$. In the meridional plane the thin disk's radial component profile could partially influence the determination of the velocity ellipsoid tilt, if the binning is done only in the vertical direction (as proposed in their paper) but not in the radial direction.

\begin{table}
\caption{Values of the thick disk velocity dispersion tensor. Absolute values of the square root of the off diagonal elements are reported.}
 \centerline {\begin{tabular}{|ccc|rrr|}
 \hline
                    &                          &                        & $[{\mbox{ }\text{km s}}^{-1}]$ &   $[{\mbox{ }\text{km s}}^{-1}]$ & $[{\mbox{ }\text{km s}}^{-1}]$ \\
 \hline
 ${\sigma _{RR}}$ & ${\sigma _{R\phi}}$    & ${\sigma _{Rz}}$     & ${59.2 \pm 4.4}$ & ${31.1 \pm 21.2}$  & ${9.3 \pm 2.3}$ \\
                    & ${\sigma _{\phi\phi}}$ & ${\sigma _{\phi z}}$ &                   & ${47.3 \pm 7.5}$  & ${3.2 \pm 7.1}$ \\
                    &                          & ${\sigma _{zz}}$     &                   &                    & ${35.9 \pm 4.1}$ \\
\hline
\end{tabular}}
\label{Tabella01}
\end{table}

\subsection{Testing the results with the Padua Galaxy model}\label{Stabilityofresults}

As indicated at the end of Section \ref{spaccotutto}, in the spirit of a Monte Carlo approach to the error analysis, we proceed by generating a set of synthetic catalogues, numbered from $j=1,...,N_{\text{cat}}$ (with $N_{\text{cat}}$ being a high number)  by assigning to the $i^{\text{th}}$ star  a random distance $d_i$ within its photometric distance error $\Delta d_i$ and a random radial velocity $v_{r,i}$ within its radial velocity error $\Delta v_{r,i}$ (no errors are assumed in the star's coordinates $(l,b)$). Then, the procedure to disentangle the stellar thin and thick population is applied for the $j^{\text{th}}$ realization of the catalogue and for all the $j=1,...,N_{\text{cat}}$. If $N_{\text{cat}}$ is sufficiently high, the stability of the error is achieved with standard statistics tools on the $N_{\text{cat}}$ realizations of the catalogue.

Nonetheless, to be sure our method is working correctly and to get more insight on its limits and potential, we apply the methodology just described to a fully analytical catalogue. Then, we apply the Padua Galaxy model \citep[e.g.,][]{1995A&A...295..655N,2002A&A...392.1129N,2006A&A...451..125V} and \citet{PhDThesis} and references therein) to generate a full synthetic Galaxy where everything is under control to test the ability to recover the properties of the data generated by the new method developed here.
In this procedure we proceed by producing mock RAVE catalogues with prescribed relations and parameters that we ultimately want to recover with the technique here developed. Initially three synthetic stellar populations, representative of the thin and thick disks and halo are generated. The stellar structure models are taken from \citet[][]{2009A&A...508..355B} and \citet{2008A&A...484..815B} from which we generate a stellar population by assuming an initial mass function (IMF) \citep[e.g.,][]{1993MNRAS.262..545K} and the star formation rate (SFR) is assumed as a free  parameter. Once the synthetic Hertzsprung-Russel diagrams with $\hat N$ stars of the stellar populations are generated and projected in the corresponding RAVE/2MASS colours and magnitude passbands, we can distribute the $\hat N$ synthetic stars representing the mixture of three stellar populations in the phase space of the MW once a model of interstellar extinction is assumed \citep[e.g., from][]{2003A&A...409..205D}. We are particularly interested in the kinematic description the stellar populations of the disks. For them we assume double exponential spatial density profiles
\begin{equation}
\rho  = \rho _{0} \left( t \right)\exp \left( { - \frac{R}
{{h_{R} \left( t \right)}} - \frac{{\left| z \right|}}
{{h_{z} \left( t \right)}}} \right),
\end{equation}
where $t$ is age of the stellar populations that we discretize for simplicity in four temporal ranges for the thin disk:
$i=1$ for $t \in \left[ {0,3} \right[{\text{Gyr}}$ with
$\rho _{0,1}  = 2.2 \times 10^8 M_ \odot  {\text{kpc}}^{ - 3}$, $h_{R,1}  = 2.90$ kpc, $h_{z,1}  = 0.20$ kpc, $i=2$ for $t \in \left[ {3,5} \right[{\text{Gyr}}$ with $\rho _{0,2}  = 1.0 \times 10^8 M_ \odot  {\text{kpc}}^{ - 3}$, $h_{R,2}  = 2.90$ kpc, $h_{z,1}  = 0.25$ kpc,
$i=3$ for $t \in \left[ {5,7} \right[{\text{Gyr}}$ with $\rho _{0,1}  = 2.1 \times 10^8 M_ \odot  {\text{kpc}}^{ - 3}$, $h_{R,1}  = 3.10$ kpc, $h_{z,1}  = 0.28$ kpc,
$i=4$ for $t \in \left[ {7,10} \right[{\text{Gyr}}$ with
$\rho _{0,1}  = 8.0 \times 10^8 M_ \odot  {\text{kpc}}^{ - 3}$, $h_{R,1}  = 3.10$ kpc, $h_{z,1}  = 0.35$ kpc. The thick disk scale parameters adopted are $\rho _{0,thick}  = 1.1 \times 10^6 M_ \odot  {\text{kpc}}^{ - 3}$, $h_{R,1}  = 2.50$ kpc, $h_{z,1}  = 1.31$ kpc and interstellar medium parameters are $\rho _{0,1}  = 1.48 \times 10^8 M_ \odot  {\text{kpc}}^{ - 3}$, $h_{R,1}  = 4.54$ kpc, $h_{z,1}  = 0.20$ kpc. To complete the density profile from which we deduce the potential that is used in the Boltzman equation to implement the kinematic description (as, e.g., in Eqn. \eqref{eqad}) we add a simple Hernquist bulge (Hernquist 1993) and logarithmic potential for the halo. We then tune a Poisson-solver to match the observational constraints on the rotation curve, i.e. the Oort functions, the local density ratio between the stellar populations, the terminal velocity for the inner Galaxy with respect to the solar radius, the total mass inside 100 kpc etc. (see appendix A in \cite{2006A&A...451..125V} for an extended description and Pasetto  2005).

It is beyond the goal of this paper to derive the structural parameters of the MW from the RAVE data. Thus, the previous values are frozen and have to be considered as guess values assumed to test our new methodology. Other literature scale parameters, shorter in the scale length of the thick disk, can be easily tested \citep[e.g.,][]{2010ApJ...712..692C, 2011ApJ...735L..46B} as suggested in relation to the alpha-enhancements \citep[e.g.,][]{2012ApJ...752...51C, 2012arXiv1206.0740B}. Once we have fixed these parameters, we fix the trend of the velocity dispersion tensor for the thin disk in the meridional plane $\sigma _{{\text{thin}}}^2 $ (see Paper II) and arbitrary values for the velocity dispersion tensor of the thick disk $\sigma _{{\text{thick}}}^2 $ with which to generate our mock catalogue and that we want to recover with our novel method.
The Padua Galaxy modelling technique produces a mock catalogue with $\hat N \geqslant N$ synthetic values for
\begin{equation}\label{mock}
\left\{ {\hat m,\hat c,\log \hat g,\hat r_{{\text{hel}}} ,l,b,\hat v_r ,{\bm{\hat \mu }}} \right\}
\end{equation}
where $\hat m$ (and $\hat c$) are magnitude (and colour) in a desired passband, and $\log_{10} \hat g$ is the logarithm of the stellar surface gravity. Considering that the cut adopted in the surface gravity in Section \ref{dataselection} lies around the turn-off of a thick disk population with $t \in \left[ {10,12} \right[{\text{Gyr}}$, we have an almost bijective relation between stellar magnitude and distance once the \citet[][]{2003A&A...409..205D} 3D extinction map is considered, thus $\hat r_{{\text{hel}}}$ is a catalogue of photometrically determined distances for the dwarf stars of the RAVE catalogue\footnote{Apart for the test performed in this section, we will adopt everywhere the photometric distances as determined in \citet{2010A&A...522A..54Z}.}. In Eqn. \eqref{mock}, when $\hat N = N$, then ${l,b}$ are exactly the observed RAVE stellar directions, $\hat v_r$ (and ${\bm{\hat \mu }}$) are the radial velocities (and eventually proper motions) for our synthetically generated catalogue.

We extensively tested the new technique against the ability to recover
\begin{itemize}
	\item the correct thick disk velocity dispersion tensor with which we generate the RAVE mock catalogues,
	\item the influence of large-scale effects on the thick disk velocity ellipsoid (see also Paper II),
	\item the minimal sky coverage for which we can correctly recover the underlying thick disk kinematics,
	\item the possibility to correctly disentangle the $v_{V, \odot } $,
	\item the role of the $\log_{10} g$ to disentangle dwarf and giant stars in our sample.
\end{itemize}
But one of the most interesting results of our extended analysis on the stability and selection cut criteria is probably the exercise presented in Appendix C, where we show that the methodology developed here mathematically can retrieve the correct thick disk velocity ellipsoid even \textit{without } the use of photometric distances.

It is possible to make further cuts, for example in the colour magnitude diagram (CMD), but as shown
in \citet{2008A&A...480..753V} or \citet{2008MNRAS.391..793S} this
drastically reduces the available number of stars.  Large numbers of stars, however, are the means of our present method to reduce the error bars. Interesting different approaches have also been discussed in the literature, e.g. by \citet{2003A&A...409..523R,1993A&AS...97..951S} or \citet{1990ApJ...357..435C,1989ApJ...339..106R}. The applicability of their methods to RAVE data is also interesting but beyond the scope of the present paper.

\subsection{Discussion of the results}\label{Discussionandconclusion}
In the following subsections we will first take a look at the assumptions and approximations trying to verify them aposteriori and then we will improve our results.

\subsubsection{Looking to the past}
\label{aposteriori}

\begin{figure}
\resizebox{\hsize}{!}{\includegraphics{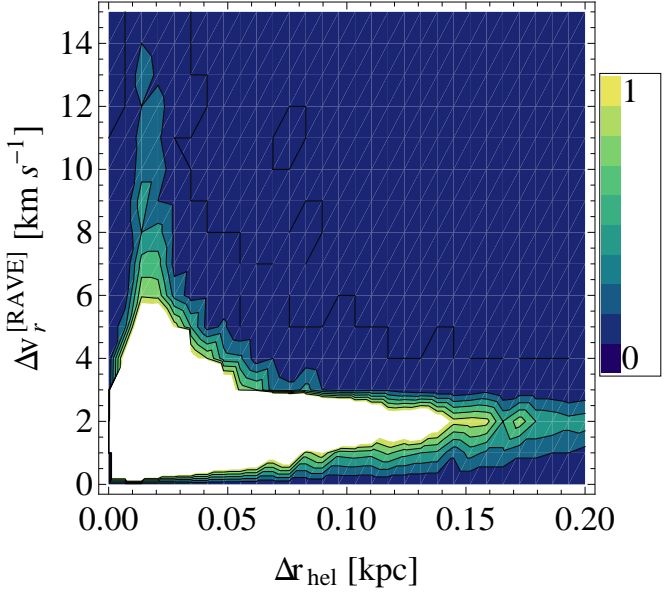}}
\caption{Plot of the errors in velocity and position: $\Delta v_r^{\text{RAVE}} $ is the error in the radial velocity as observed in the RAVE survey.  $\Delta r_{\rm{hel}} $ is the error in the photometric distances as inferred in \cite{2010A&A...522A..54Z}. The binning is 0.5 km s$^{ - 1}$ and 50 pc with a scale colour normalized to the highest number of stars per bin (in light yellow) to 0 stars per bin (dark blue contour).}
\label{Figure05}
\end{figure}

Up to now, we approached the problem of the determination of the thick disk mean velocity and its velocity dispersion tensor. This approach permitted us to determine the kinematic properties of the thick disk component in the framework of an isothermal velocity ellipsoid model (Table \ref{Tabella01}). Nevertheless, in the literature the ellipsoid is commonly described as a velocity distribution independent of Galactic position because of the difficulties in achieving a more accurate description. However, both these hypotheses are without a rigorous theoretical basis for the thick disk component of our Galaxy.
Here, thanks to the RAVE survey, we can investigate this aspect further.

In Figure \ref{Figure05} we plot the errors in the velocity and position for the sample of stars in the mixture of thin and thick disks as obtained from the catalogue (same stars as in Figure \ref{SkyDist}). As already pointed out in Section \ref{RadialvelocitycomponentfromtheGalaxydifferentialrotation}, we work with the stars with larger errors in the velocity. Thus the bulk of the sample has radial velocity uncertainties up to ${\text{6 km s}}^{ - 1} $, slightly higher than the average error produced by the RAVE survey \citep[e.g.,][]{2006AJ....132.1645S}. The photometric distance errors are typically of the order of 0.15 kpc for our selected sample \citep[][]{2010A&A...522A..54Z}.

A free parameter of our modelling approach is the mean
azimuthal velocity of the RAVE sample we select. This quantity is related to the asymmetric drift of the stellar populations, the rotational lag of the thick disk and the circular velocity of rotation. The thick disk subsample of
the RAVE catalogue analysed here appears to lag behind the Solar LSR by $\cong 49
{\mbox{ }\text{km s}}^{-1}$. The general behaviour of the asymmetric drift is not known a priori, especially in its vertical trend and for a
mixed set of thin and thick disk populations. For a single stellar
population in the disk, the mean circular streaming velocity can
theoretically be predicted, starting from the Jeans equation in cylindrical
symmetry. We can approximate the term ${\mathbf{\bar v}}_c \left( {\mathbf{x}} \right)$ in Eqn. \eqref{EQ01} as
\begin{equation}
{\mathbf{\bar v}}_c \left( {\mathbf{x}} \right) = {\mathbf{\bar v}}_c \left( {R,z} \right) = \left( \begin{array}{l}
 \left| {{\mathbf{\bar v}}_c \left( {R,z} \right)} \right|\frac{{r_{{\rm{hel}}} \cos b\sin l}}{R} \\
 \left| {{\mathbf{\bar v}}_c \left( {R,z} \right)} \right|\frac{{R_ \odot   - r_{{\rm{hel}}} \cos b\cos l}}{R} \\
 0 \\
 \end{array} \right)
\end{equation}
with
\begin{equation}\label{eqad}
\begin{array}{c}
 \left| {{\bf{\bar v}}_{\rm{c}} \left( {R,z} \right)} \right| = \left[ {V_{{\rm{LSR}}}^2  - \frac{{\partial \ln \rho }}{{\partial \ln R}}\left( {\sigma _{{\rm{RR}}}^2  + \sigma _{{\rm{Rz}}}^2 } \right) + \left( {\sigma _{{\rm{RR}}}^2  + \sigma _{\varphi \varphi }^2 } \right)} \right. \\
  + \left. {R\left( {\frac{{\partial \sigma _{{\rm{RR}}}^2 }}{{\partial R}} + \frac{{\partial \sigma _{{\rm{Rz}}}^2 }}{{\partial z}} + \frac{{\partial \Phi _{{\rm{tot}}} \left( {R,z} \right)}}{{\partial R}}} \right)} \right]^{1/2}  \\
 \end{array}
\end{equation}
where ${{\Phi }_{\text{Tot}}}$ is the total potential of the MW. This indicates
how the terms of the velocity dispersion profiles, the circular velocity
and the density profile act together to play a crucial role in the
prediction of the streaming circular velocity and in its axisymmetric
approximation. Assuming that the selected data sample is mostly representative of the thick
disk component, the previous Eqn. \eqref{eqad} is reduced, to:
\[\sqrt{\left( \frac{\partial \ln \rho \left( {{R}_{\odot }},0 \right)}{\partial \ln R}+1 \right)\sigma _{RR}^{2}-\sigma _{\phi \phi }^{2}+v_{c}^{2}}\cong 201{\mbox{ }{\text{km s}}^{ - 1}} \]
where ${{h}_{R}}\cong 2.9{\text{kpc}}$ is the adopted value for the scale length of
a double exponential density profile, e.g., \citet{2004ASPC..317..203V}.
${{v}_{c}}$ is the circular velocity of the Galaxy in the solar
neighbourhood as above, and $\sigma _{RR}^{2}$ and $\sigma _{\phi \phi
}^{2}$ take the values derived in our paper.  The expected value for a pure
thick disk component is not so far from the best fit value derived in our
study, namely $\left\| {{{\mathbf{\bar{v}}}}_{c}}
\right\|=179{\mbox{ }\text{km s}}^{-1}$. We point out here that this
permits us to impose only an \textit{upper limit} on the expected value of
the streaming circular velocity for the mixed sample of thin and thick
disks we analysed. This is thus not necessarily in disagreement with previous
studies predicting lower values \citep[e.g.,][]{2003A&A...398..141S} because
of the thin disk influence. A further cut in the latitude can in principle
reconcile these values but unfortunately reduces the number of stars that
we can retain with our iterative procedure and consequently increases greatly
the error bars in the velocity dispersion tensor.

Thus we prefer to consider $\left\| {{\mathbf{v}}_{c}} \right\|$ as a free
tuning parameter in our approach, remembering that if the thick disk
results from a sudden heating due to an infalling satellite, it can be
kinematically more decoupled from other components of the MW \citep[e.g.,][]{1986ApJ...309..472Q, 1996ApJ...460..121W,  2010A&A...510L...4S, 2011ApJ...738....4B}. The overlap between rotational velocity and peculiar motion of the Sun relative to the LSR can not be disentangled with the use of an isothermal, non-local sample of stars so it remains beyond the scope of the present work \citep[see e.g.,][]{2010MNRAS.tmp..149S}.

\begin{table}\label{RableRef}
\caption{Overview of the data sets and type of analysis performed by recent literature works (not based on RAVE data). Column one refers to the number of stars effectively used in the analysis of the thick/thin disk, not to the total number of stars produced by the database. HR/LR in column two stands for high/low spectral resolution respectively as specified in the papers of column  three. For comparison we report the number of stars used in the present study in the last row. }
\begin{tabular}{|c|l|c|}
 \hline
  no. stars & Type of analysis                    & References \\
 \hline
  451     & Chemistry (HR spectra), kinematics  & (1)    \\
  176 		 & Chemistry (HR spectra), kinematics & (2)    \\
  250 	 	 & Chemistry (HR spectra), kinematics & (3)    \\
 1,498    & Strömgren photometry, kinematics    &     (4)    \\
  306     & Chemistry (HR spectra), kinematics  &     (5)(2)(9)(11)   \\
  412     & Kinematics (HR spectra)             &     (6)(7)(12) \\
   76     & Chemistry (HR spectra), kinematics &     (8)    \\
   102    & Chemistry (HR spectra), kinematics & (9)    \\
17,277    & Chemistry (LR spectra), kinematics &  (10)    \\
23,767    & Chemistry (LR spectra), kinematics &  (13)(14) \\
 \hline
38,805    & Kinematics (HR spectra) &  present work \\
 \hline
\end{tabular}
\tablebib{
(1)~\citet{2009A&A...497..563N}; (2) \citet{2006MNRAS.367.1329R}; (3) \citet{2004AN....325....3F}; (4) \citet{2011A&A...530A.138C};
(5) \citet{2011MNRAS.412.1203N}; (6) \citet{2012ApJ...747..101M}; (7) \citet{2012ApJ...751...30M}; (8) \citet{2012AJ....144...20A};
(9) \citet{2005A&A...433..185B}; (10) \citet{2011ApJ...738..187L}; (11) \citet{2010A&A...511L..10N};
(12) \citet{2006AJ....132.1768G}; (13) \citet{2012ApJ...751..131B}; (14) \citet{2012arXiv1202.2819B}.}
\end{table}

\subsubsection{Looking to the future}
\label{futuroeoltre}

In principle, it is nowadays possible to disentangle the thick and  thin disks using  selection criteria based on the chemical properties of the stars \citep[e.g., see the recent review by][]{2011arXiv1109.4010N}.  In brief, \citet[][]{2009A&A...497..563N} determined the trend of abundance ratios as a function of $\left[ {\text{Fe}/\text{H}} \right]$ from 451 high-resolution spectra of F, G, and K main-sequence stars in the solar neighbourhood confirming the long known bimodal distribution of  the $\left[ {\alpha /\text{Fe}} \right]$ ratio for the disk stars (with  the thin disk stars  less alpha-enhanced than the thick disk ones), see for instance   \citet[][]{2006MNRAS.367.1329R} or \citet{2004AN....325....3F} and references therein.
Along the same line of thought, \citet{2011A&A...530A.138C}  determined the ratios $\left[ {\text{Fe}/\text{H}} \right]$ and $\left[ {\alpha /\text{Fe}} \right]$ for 1498 selected stars of the Geneva-Copenhagen Survey from Str$\ddot{\text{o}}$mgren photometry,  even if their data do not provide a clear bimodal distribution between thin and thick disk stars as far as  the $\left[ {\alpha /\text{Fe}} \right]$ ratio is concerned.
\citet{2011MNRAS.412.1203N} investigated the separation of thin and thick disks using a  combined index of $\left[ {\text{Fe}/\text{H}} \right]$, $\left[ {\alpha /\text{Fe}} \right]$ and the heavy element Eu for a  sample of 306 stars.
\citet[][]{2011ApJ...737....9R} studied the vertical and radial gradients in metallicity and alpha-elements  for a sample of selected thick disk stars.
\citet{2012ApJ...747..101M,2012ApJ...751...30M} studied 412 red giant stars in the direction towards the South Galactic Pole  trying to describe kinematics, chemistry,  and content of dark matter  in the MW disks \citep[but see e.g.,][]{2012arXiv1205.4033B}. Finally,
\citet{2012AJ....144...20A} conducting a detailed abundance analysis and atmospheric parameters of 76 stars in the thin and thick disks with very high-resolution spectra ($\Re\simeq60,000$)  reaching conclusions similar to those found by other authors.

Common to all the above studies based on chemical properties (see Table \ref{RableRef}), is the small  number of stars that are considered and the limited spatial coverage in about the solar neighbourhood. In contrast, our method, owing to its statistical nature, first requires a high number of stars such as that produced by RAVE, and second does not need chemical information to work properly. It is worth recalling that data on radial velocities (from medium resolution spectroscopy from which the kinematics is derived) and photometry are easier to gather than good chemical abundances (from high resolution spectroscopy). Our approach can easily be applied to large sets of spectroscopic data such as the continuously growing RAVE survey (or Gaia in the near future). Finally, our method makes use only of kinematic data, leaving the coupling between kinematics and chemistry to be investigated in a forthcoming paper \citep[][]{CorradoP} %This is typical of the data provided by  SEGUE  \citep[e.g.,][]{2011ApJ...738..187L, 2012ApJ...751..131B, 2012arXiv1202.2819B} which are based on low resolution spectra and hence chemical data of lower quality. }

Although in recent years we have seen continuous improvements in Galactic modelling, see e.g. the Besançon model \citep[][]{2003A&A...409..523R}, the Padova Galaxy model (used in this paper), DF based models \citep[e.g.,][]{2012arXiv1207.4917B} or the Galaxia model \citep{2011ApJ...730....3S}, we still lack a good theoretical framework coupling kinematics and dynamics with population synthesis and chemistry \citep[e.g.,][]{2009MNRAS.396..203S,2011MNRAS.411.2586J, 2010MNRAS.402..461J}. 

Neglecting the radial and vertical dependence for the \textit{ thick disk} velocity ellipsoid is an almost universal assumption that indeed produces acceptable results for the vertical tilt in our case as well. Nevertheless, we can further explore this working hypothesis by making use of the available photometric distances.
Although the proximity of the stars selected grants the validity of our approximations, it limits the exploration of the more distant zones of the thick disk. For every sample statistically representative of the thick disk population that we obtain with the previously outlined method, we split the data set into two regions: within and beyond the Sun's Galactocentric position, $R \leqslant R_ \odot  $
 and $R > R_ \odot  $ respectively (where $R$ is the cylindrical Galactocentric radius).
The values obtained for the velocity dispersion tensor are then statistically averaged, as done to obtain Table \ref{Tabella01}, and are listed in Tables \ref{Tabella02} and \ref{Tabella03}.

\begin{table}
\caption{Values of the thick disk velocity dispersion tensor for $R<R_{\odot}$. Units as in Table \ref{Tabella01}}
 \centerline {\begin{tabular}{|ccc|ccc|}
 \hline
                    &                          &                        & $[{\mbox{ }\text{km s}}^{-1}]$ &   $[{\mbox{ }\text{km s}}^{-1}]$ & $[{\mbox{ }\text{km s}}^{-1}]$ \\
 \hline
 ${\sigma _{RR}}$ & ${\sigma _{R\phi}}$    & ${\sigma _{Rz}}$     & ${60.2 \pm 7.1}$ & ${37.6 \pm 21.7}$  & ${13.3 \pm 9.8}$ \\
                    & ${\sigma _{\phi\phi}}$ & ${\sigma _{\phi z}}$ &                   & ${44.7 \pm 8.1}$  & ${4.0 \pm 7.2}$ \\
                    &                          & ${\sigma _{zz}}$     &                   &                    & ${37.2 \pm 5.7}$ \\
\hline
\end{tabular}}
\label{Tabella02}
\end{table}
\begin{table}
\caption{Values of the thick disk velocity dispersion tensor for $R>R_{\odot}$.}
 \centerline {\begin{tabular}{|ccc|ccc|}
 \hline
                    &                          &                        & $[{\mbox{ }\text{km s}}^{-1}]$ &   $[{\mbox{ }\text{km s}}^{-1}]$ & $[{\mbox{ }\text{km s}}^{-1}]$ \\
 \hline
 ${\sigma _{RR}}$ & ${\sigma _{R\phi}}$    & ${\sigma _{Rz}}$     & ${55.8 \pm 6.5}$ & ${35.5 \pm 20.2}$  & ${9.6 \pm 7.6}$ \\
                    & ${\sigma _{\phi\phi}}$ & ${\sigma _{\phi z}}$ &                   & ${45.2 \pm 7.3}$  & ${3.8 \pm 3.1}$ \\
                    &                          & ${\sigma _{zz}}$     &                   &                    & ${36.3 \pm 4.1}$ \\
\hline
\end{tabular}}
\label{Tabella03}
\end{table}

The entries of Tables \ref{Tabella02} and \ref{Tabella03} hardly show any difference between the stars outside and inside the solar circle. Nonetheless, the suspicion arises that within the errors, there may be some evidence for a general increase in the velocity dispersion of the thick disk along  the radial direction (the velocity dispersion is higher for star inside the solar circle than outside). This implies that  for the   thick disk alone  the classical isothermal picture needs to be improved. The thick disk velocity dispersion tensor seems to imply a dependence on the position in the meridional plane $\sigma _{ij}^{\text{thick}}  = \sigma _{ij}^{\text{thick}} \left( {R,z} \right)$. The effect is, however,  small due to the small range of distances sampled with our data. We have higher error bars in the inner sample because of the cut we have applied on the radial velocity errors  (see Eqn.  \eqref{EQ07}) that retains a higher number of stars in the anti-centre direction (see Figure \ref{SkyDist}). As a consequence, despite the stellar density decrease towards the outer regions of the Galaxy, the error bars of the inner and outer samples of stars  are comparable in size.

This suggests a way of evaluating  the effect of the distance if a gradient in the vertical dependence of the absolute value of the azimuthal  velocity occurs along the thick disk \citep[e.g.,][]{2000AJ....119.2843C}.
Eqn. \eqref{eqad}, which holds for small distances from the plane \citep[e.g.,][their Eqns. (31) to (33)]{2006A&A...451..125V} can be used to create mock catalogues with radial and vertical gradients in $\left| {{\bf{\bar v}}^i_{\rm{c}} \left( {R,z} \right)} \right|$ where $i$ refers to the thin and thick disk once the populations are embedded in a common potential $\Phi _{{\rm{tot}}}$. In this way we can check the role of small gradients, if any, in $\sigma _{ij}^{{\text{thick}}} = \sigma _{ij}^{{\text{thick}}}\left( {R,z} \right)$ on $\left| {{\bf{\bar v}}^\text{thick}_{\rm{c}} \left( {R,z} \right)} \right|$ or we can simply artificially place a gradient in $\left| {{\bf{\bar v}}^\text{thick}_{\rm{c}} \left( {R,z} \right)} \right|$.
This approach has been tested with Eqn.\ \eqref{eqad} and gradients up to 30 ${\text{km s}}^{-1} {\text{kpc}}^{-1}$. The resulting values for the thick disk velocity ellipsoid are within the error bars presented in Table \ref{Tabella01} \citep[see also][]{2003ASPC..298..153B}. On the one hand, this is because of the limited vertical and radial extension reached by  our sample, which reduces large scale effects.  On the other hand, this happens  due to the method itself not involving a selection based on chemical abundances. For example, a  vertical gradient in mean azimuthal velocity  of about 20 ${\text{km s}}^{-1} {\text{kpc}}^{-1}$ was found by \citet{2010A&A...510L...4S} even though their result refers to $\left| z \right| \in \left] {1,3} \right[{\text{kpc}}$, i.e., a region beyond the range spanned by our sub-sample of the RAVE catalogue.

The different result in the work of \citet{2008A&A...480..753V} is due to
their data sample being based on fewer than 600 stars in the direction of
the south Galactic pole (SGP). This leads to a higher value of ${{\sigma
}_{RR}}$, on which the influence of halo stars is yet to be investigated
\cite[see, e.g.,][]{2000AJ....119.2843C,2009ApJ...698.1110S}.  Particular
attention should be paid to the interpretation of the components that are
mixed with the azimuthal component: ${{\sigma }_{R\phi }}=29.4 \pm
17.2{\mbox{ }{\text{km s}}^{ - 1}} $, ${{\sigma }_{\phi z}}=5.8 \pm
5.1\ {\mbox{ }{\text{km s}}^{ - 1}} $.  The first is reminiscent of
the well-known vertex deviation. For the thin disk the vertex deviation can
be predicted from ${{\sigma }_{R\phi }}$ and amounts to an angle of
$\cong 21{}^\circ $. Its large error is probably due to the thick disk
components closer to the MW plane which could induce circular velocity gradients that we were unable to detect with our method. The ${{\sigma }_{\phi z}}$ component is
a further indication of the coupling of the vertical and azimuthal velocity
components.

Finally we point out that the method we have developed can be extended to an arbitrary number of populations provided that first the formalism is expanded to include higher order cumulants, second each population is suitably sampled (sufficiently high number of stars), and third each population is characterized by at least one distinct kinematical parameter. In our case, we present evidence of the existence of at least two distinct populations whose characterizing parameters are the second order velocity dispersion tensors.
To prove this statement we perform the following experiment. Suppose that the solar position is moved  closer towards the Galactic centre, for instance at the position ${R_{\hat  \odot }} \leqslant  5$ kpc. Then we set up a mock  catalogue   centred on the new position of the Sun, ${\hat R_\odot }$. By construction, the velocity dispersions of the thin and thick disk stars are the same (no correction for different extinction is applied). Since the kinematical parameters of the populations are identical, the method fails to converge to a solution.  This result can be better understood when looking at  Fig.\ref{Figure000}, where now we imagine that the Sun  is located at ${\hat  R_\odot }$ and the vectors representing the velocity distribution of thick disk stars (light blue arrows) have the same length of  those for the  thin disk stars (olive-green arrows).  Therefore, there is no kinematic way of distinguishing two different populations from the radial velocity distribution of the mixture (the thick green arrows). However, this does imply that the two populations could be separated by  considering other parameters  such as  chemistry, alpha-enhancements etc.

\section{Discussion and conclusions}\label{thend}
 In the sample extracted from the RAVE catalogue that we analysed,  two populations of stars with different kinematics are found that correspond to the thin and thick disks, even though our analysis actually focused only on the thick disk.

The major difference with respect to previous studies in  the literature is the peculiar spatial volume covered by RAVE, which allows us to determine the kinematics of the thick disk  not by extrapolating from a single small field-of-view  to the whole thick disk, but by directly measuring the kinematics by studying the much larger volume of the thick disk observed by RAVE (see also Paper II).
Over this extension, the key thick disk parameters are determined as:
\begin{enumerate}
	\item two components of the solar motion relative to the solar LSR, namely as $ v_{U, \odot }  = \left( {9.87 \pm 0.37} \right){\text{km s}}^{ - 1} $ and $v_{W, \odot }  = \left( {8.01 \pm 0.29} \right){\text{km s}}^{ - 1}$,
	\item the rotational lag of the thick disk component relative to the LSR $	{{v}_{\text{thick,lag}}}\cong 49\pm6{\mbox{ }\text{km s}}^{-1}$,
	\item the velocity dispersion tensor of the thick disk considered to be an isothermal population: $\sigma _{RR}^{}  = (56.1 \pm 3.8) {\text{km s}}^{ - 1}$, $\sigma _{R\varphi }  = (29.4 \pm 17.2) {\text{km s}}^{ - 1}$, $\sigma _{Rz}  = (10.1 \pm 3.3) {\text{km s}}^{ - 1}$, $\sigma _{\varphi \varphi }  = (46.1 \pm 6.7) {\text{km s}}^{ - 1}$, $\sigma _{\varphi z}  = (5.8 \pm 5.1) {\text{km s}}^{ - 1}$, $\sigma _{zz}  = (35.1 \pm 3.4) {\text{km s}}^{ - 1}$.
\end{enumerate}
Moreover, we mention that the missing full determination of the Sun's velocity vector relative to the LSR is just a choice. This does not mean at all the impossibility for RAVE to characterize completely the solar neighbourhood \citep[see, e.g.,][]{2010MNRAS.tmp.1636K,2011MNRAS.412.1237C}. We defer the study of the thin disk kinematics to a companion paper (Paper II).

The determination of the thick disk velocity dispersion tensor and its behaviour in the $(O;R,z)$ plane is just a small step in the investigation of this Galactic component. The presence of a small gradient, especially in the $\sigma_{RR}$ component, does not represent the failure of the classical picture of an isothermal description within the range of distances investigated  but it is an example of the quality of data that a radial velocity survey like RAVE can provide to confirm, extend or investigate new ideas.

The thick disk is a prominent feature of our galaxy \citep{1983MNRAS.202.1025G,1982PASJ...34..365Y,2008ApJ...673..864J,2009AJ....137.4377Y} and of external disk galaxies \citep[e.g.,][]{2011ARA&A..49..301V, 2006AJ....131..226Y}. Several possible formation mechanisms have been suggested for the MW thick disk formation. For instance it may be related to the influence of massive satellites that can either heat pre-existing disks or contribute by being accreted \citep[e.g.,][]{1986ApJ...309..472Q, 1993ApJ...403...74Q,2003ApJ...597...21A}.
\cite{1996ApJ...460..121W} showed in detail how low-mass satellites, while rapidly sinking into the potential well of a galaxy, could substantially heat a disk. \citet{2004ApJ...612..894B} investigated the influence of gas-rich mergers. Observational evidence of this process is presented by \cite{1996A&A...305..125R}, \cite{1996IAUS..169..681R}, \cite{2002ApJ...574L..39G} and \cite{2006ApJ...639L..13W}. After the merger, it is plausible that the star formation stopped for a while until the gas assembled again in the thin disk \citep[e.g., see extensive discussions in][]{2002ARA&A..40..487F,2002EAS.....2..295W}. Finally the MW can produce thick disk features in itself by radial migration processes \citep{2008ApJ...675L..65R, 2009MNRAS.396..203S} or from disruption of massive star clusters \citep[e.g.,][]{2002MNRAS.330..707K}.

In order to distinguish the role of these different scenarios the improved kinematics data expected from the forthcoming astrometric Gaia satellite are fundamental \citep[e.g.,][]{2012A&A...543A.100R}, especially in order to constrain time-evolving self-consistent dynamical and chemical models \citep[e.g.,][]{2011MNRAS.415.1469R, 2011ApJ...737....8L, 2011MNRAS.415.2652H, 2012ApJ...747..101M}.
If the thick disk is formed from accreted stars, e.g. during a merger event, then no vertical gradient, $\left[ {\text{Fe}/\text{H}} \right]\left( z \right)$  is expected for the thick disk \citep[but see][]{2011A&A...525A..90K}. The radial mixing for a MW in isolation \citep[e.g.][]{2009MNRAS.396..203S} is probably not an efficient mechanism to remove the metallicity (or alpha elements) radial gradients \citep[if any, see e.g.,][]{2011ApJ...737....9R} beyond $R>9$ kpc because of the steeply decreasing  probability of radial migration suggested by \citet[][]{2011ApJ...735L..46B}. Radial migration remains still a mechanism to be theoretically defined beyond the mere N-body numerical experiment \citep[][]{2012MNRAS.422.1363S, 2012arXiv1205.6475M}, while mergers can sensibly flatten the migration probability or also anti-correlate it with the radial density profile \citep[][]{2012MNRAS.420..913B}. A correlation between $\left\| {{\mathbf{\bar v}}_c } \right\|$ and $\left[ {\text{Fe}/\text{H}} \right]$ would probably disfavour a migration scenario and slow heating mechanisms \citep[e.g.,][]{2011MNRAS.412.1203N}.

\begin{acknowledgements}
We acknowledge the referee for the constructive report. S.P. wants to thank B. Fuchs, A. Just and J. Binney for comments on the technique and results of the paper and P. Re Fiorentin, S. Jin and I. Minchev for careful reading of the manuscript.
Numerical computations have been partially performed with supercomputers at the John von Neumann - Institut f$\ddot{u}$r Computing (NIC) - Germany  (NIC-project number 2979). We acknowledge partial funding from Sonderforschungsbereich SFB 881 ``The Milky Way System'' (subprojects A5 and A6) of the German Research Foundation (DFG).
Funding for RAVE has been provided by: the Australian Astronomical
Observatory; the Leibniz-Institut fuer Astrophysik Potsdam (AIP); the
Australian National University; the Australian Research Council; the
French National Research Agency; the German Research Foundation (SPP 1177 and SFB 881);
the European Research Council (ERC-StG 240271 Galactica); the Istituto Nazionale di
Astrofisica at Padova; The Johns Hopkins University; the National Science Foundation of
the USA (AST-0908326); the W. M. Keck foundation; the Macquarie University; the
Netherlands Research School for Astronomy; the Natural Sciences and Engineering Research
Council of Canada; the Slovenian Research Agency; the Swiss National Science Foundation;
the Science \& Technology Facilities Council of the UK; Opticon; Strasbourg Observatory;
and the Universities of Groningen, Heidelberg and Sydney. The RAVE web site is at
http://www.rave-survey.org.
\end{acknowledgements}

\bibliographystyle{aa}
\bibliography{BiblioArt}

\appendix
\onecolumn
\section{The whole sky symmetry}
\begin{figure*}
\sidecaption
\includegraphics[width=10cm]{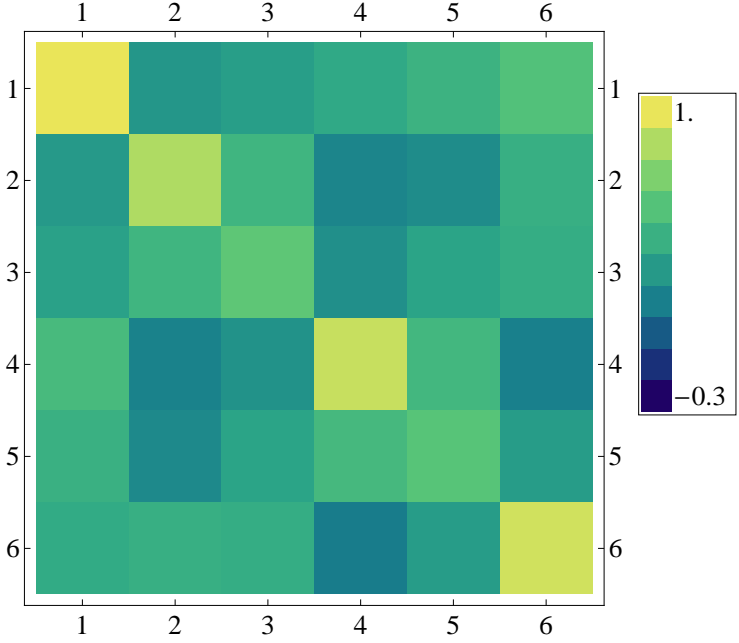}
\caption{The inverse matrix of the Eqn.\ \eqref{EQ13} for the best fit
model. The matrix closely resembles the fully
analytical case except for the numerical part.
Different shades of colour are applied to visualize  the symmetries.}
\label{Figure03}
\end{figure*}

Here we outline the procedure to pass from a radial velocity set of data to the true velocity first order moments in the case of an all-sky-survey coverage. This is an application of what is shown in Section \ref{sguardo} that we used as reference case. In this case the
matrix in Eqn. \eqref{EQ13} can be computed directly using spherical coordinates
$\left\{ {\hat x,\hat y,\hat z} \right\} = \left\{ {\cos b\cos l,\cos b\sin
l,\sin b} \right\}$. The generic element of the matrix, ${\mathbf{\bar
M}}_{\left[ 2 \right]} $,
\begin{equation}
\left( {\mathbf{\bar M}}_{\left[ 2 \right]} \right)_{i,j}  = \left( {\begin{array}{*{20}c}
   {\bar M_{\left[ 2 \right]1,1} } &  \ldots  & {\bar M_{\left[ 2 \right]1,6} }  \\
    \vdots  &  \ddots  &  \vdots   \\
   {\bar M_{\left[ 2 \right]1,6} } &  \cdots  & {\bar M_{\left[ 2 \right]6,6} }  \\
 \end{array} } \right)
\label{EQA01}
\end{equation}
can be computed as
\begin{equation}
{{\mathbf{\bar M}}_{\left[ 2 \right]}} = \frac{1}
{{4\pi }}\int_{S^2 }^{} {\bar M_{\left[ 2 \right]i,j} d\Omega }
\label{EQA02}
\end{equation}
where $S^2  = \left[ {0,2\pi } \right[ \times \left[ { - \pi /2,\pi /2}
\right[$ and the solid angle $d\Omega  = dld\left( {\cos b} \right)$.  Or,
element by element:
\begin{equation}
\left( {\begin{array}{*{20}c}
   {\bar M_{\left[ 2 \right]1,1} } &  \ldots  & {\bar M_{\left[ 2 \right]1,6} }  \\
    \vdots  &  \ddots  &  \vdots   \\
   {\bar M_{\left[ 2 \right]1,6} } &  \cdots  & {\bar M_{\left[ 2 \right]6,6} }  \\
 \end{array} } \right) = \frac{1}
{5}.\left( {\begin{array}{*{20}c}
   1 & 0 & 0 & {\frac{1}
{3}} & 0 & {\frac{1}
{3}}  \\
   0 & {\frac{2}
{3}} & 0 & 0 & 0 & 0  \\
   0 & 0 & {\frac{2}
{3}} & 0 & 0 & 0  \\
   {\frac{1}
{3}} & 0 & 0 & 1 & 0 & {\frac{1}
{3}}  \\
   0 & 0 & 0 & 0 & {\frac{2}
{3}} & 0  \\
   {\frac{1}
{3}} & 0 & 0 & {\frac{1}
{3}} & 0 & 1  \\

 \end{array} } \right).
\label{EQA03}
\end{equation}
From its inverse we can easily obtain the elements of the velocity
ellipsoid as:
\begin{equation}
{\bm{\sigma }}^2  \equiv \left( {\begin{array}{*{20}c}
   {\sigma _{RR}^2 }  \\
   {\sigma _{R\phi }^2 }  \\
   {\sigma _{Rz}^2 }  \\
   {\sigma _{\phi \phi }^2 }  \\
   {\sigma _{\phi z}^2 }  \\
   {\sigma _{zz}^2 }  \\

 \end{array} } \right) = \frac{3}
{2}.\left( {\begin{array}{*{20}c}
   {4\sigma _{\parallel RR}^2  - \sigma _{\parallel \phi \phi }^2  - \sigma _{\parallel zz}^2 }  \\
   {5\sigma _{\parallel R\phi }^2 }  \\
   {5\sigma _{\parallel Rz}^2 }  \\
   {4\sigma _{\parallel \phi \phi }^2  - \sigma _{\parallel RR}^2  - \sigma _{\parallel zz}^2 }  \\
   {5\sigma _{\parallel \phi z}^2 }  \\
   {4\sigma _{\parallel zz}^2  - \sigma _{\parallel RR}^2  - \sigma _{\parallel \phi \phi }^2 }  \\

 \end{array} } \right).
\label{EQA04}
\end{equation}
In general, all this is valid when the data have a spherically symmetric distribution.
With real data, owing to the partial sky coverage, the matrix \eqref{EQA03} may substantially differ from the  symmetric case. To take this into account,  we use the matrix \eqref{EQA03} as a mask, i.e. as a constraint on the relative weight that the generic matrix element ${\bar M_{[2]i,j} }$ has with respect to another element ${\bar M_{[2]k,l} }$ when trying to maximize the parameter $q$ of Eqn. \eqref{EQq}, i.e. to control the coupling of the off-diagonal blocks of the operator matrix ${\mathbf{P}}$. The generic matrix element to be determined  requires integrals of many complex trigonometric functions, thus implying long tedious calculations. Fortunately, many elements are null by symmetry.  For instance, because there exist only three linearly independent isotropic fourth-rank tensors related to the Kronecker delta tensor by ${\mathcal{I}_{ijkl}} = {\delta _{il}}{\delta _{ik}}$, ${\mathcal{J}_{ijkl}} = {\delta _{ik}}{\delta _{jl}}$ and ${\mathcal{K}_{ijkl}} = {\delta _{ij}}{\delta _{kl}}$,   the generic symmetric fourth-rank isotropic tensor ${X_{[4]}}$ can be expressed as a linear  combination of these (e.g., with coefficient ${X^s},{X^a},{X^{Tr}}$):
\begin{equation}\label{lll}
{X_{[4]}} = {X^s}\left( {\frac{1}{2}\left( {\mathcal{I} + \mathcal{J}} \right) - \frac{1}{3}\mathcal{K}} \right) + {X^a}\left( {\mathcal{I} - \mathcal{J}} \right) + {X^{Tr}}\frac{1}{3}\mathcal{K},
\end{equation}
so that only  the different terms can be easily singled out  from Eqn. \ref{lll}. Comparing the shape of the matrix in Eqn. \eqref{EQA03} with the true case of RAVE data from Fig. \ref{Figure03} shows that in most cases nearly symmetric conditions apply.  Indeed,  matrices in Eqn.\ \eqref{EQ13}, ${\mathbf{\bar M}}_{\left[
2 \right]} $, Eqn. \eqref{EQ13III}, ${\mathbf{\bar M}}_{\left[ 3 \right]} $,
and \eqref{EQ13IV}, ${\mathbf{\bar M}}_{\left[ 4 \right]} $  closely resemble
the case of spherical symmetry (see, e.g., Fig.\
\ref{Figure03} for the matrix of Eqn. \eqref{EQ13}).
For the third moment we have
\begin{equation}
{{\mathbf{\bar M}}_{\left[ 3 \right]}} = \frac{1}
{{4\pi }}\int_{S^2 }^{} {\bar M_{\left[ 3 \right]i,j} d\Omega } \label{EQA05}
\end{equation}
to get
	\begin{equation}
\left( {\begin{array}{*{20}c}
   {\bar M_{\left[ 3 \right]1,1} } &  \ldots  & {\bar M_{\left[ 3 \right]1,10} }  \\
    \vdots  &  \ddots  &  \vdots   \\
   {\bar M_{\left[ 3 \right]1,10} } &  \cdots  & {\bar M_{\left[ 3 \right]10,10} }  \\

 \end{array} } \right) = \frac{1}
{7} \cdot \left( {\begin{array}{*{20}c}
   1 & 0 & 0 & {\frac{2}
{5}} & 0 & {\frac{2}
{5}} & 0 & 0 & 0 & 0  \\
   0 & {\frac{3}
{5}} & 0 & 0 & 0 & 0 & {\frac{2}
{5}} & 0 & {\frac{1}
{5}} & 0  \\
   0 & 0 & {\frac{3}
{5}} & 0 & 0 & 0 & 0 & {\frac{1}
{5}} & 0 & {\frac{2}
{5}}  \\
   {\frac{2}
{5}} & 0 & 0 & {\frac{3}
{5}} & 0 & {\frac{1}
{5}} & 0 & 0 & 0 & 0  \\
   0 & 0 & 0 & 0 & {\frac{2}
{5}} & 0 & 0 & 0 & 0 & 0  \\
   {\frac{2}
{5}} & 0 & 0 & {\frac{1}
{5}} & 0 & {\frac{3}
{5}} & 0 & 0 & 0 & 0  \\
   0 & {\frac{2}
{5}} & 0 & 0 & 0 & 0 & 1 & 0 & {\frac{2}
{5}} & 0  \\
   0 & 0 & {\frac{1}
{5}} & 0 & 0 & 0 & 0 & {\frac{3}
{5}} & 0 & {\frac{2}
{5}}  \\
   0 & {\frac{1}
{5}} & 0 & 0 & 0 & 0 & {\frac{2}
{5}} & 0 & {\frac{3}
{5}} & 0  \\
   0 & 0 & {\frac{2}
{5}} & 0 & 0 & 0 & 0 & {\frac{2}
{5}} & 0 & 1  \\

 \end{array} } \right),
\label{EQA06}
\end{equation}

and from its inverse
	\begin{equation}
{\mathbf{S}} \equiv \left( {\begin{array}{*{20}c}
   {S_{RRR} }  \\
   {S_{RR\phi } }  \\
   {S_{RRz} }  \\
   {S_{R\phi \phi } }  \\
   {S_{R\phi z} }  \\
   \begin{gathered}
  S_{Rzz}  \hfill \\
  S_{\phi \phi \phi }  \hfill \\
  S_{\phi \phi z}  \hfill \\
  S_{\phi zz}  \hfill \\
  S_{zzz}  \hfill \\
\end{gathered}   \\
 \end{array} } \right) = \frac{{35}}
{{24}}.\left( {\begin{array}{*{20}c}
   {4\left( {2S_{\parallel RRR}  - S_{\parallel R\phi \phi }  - S_{\parallel Rzz} } \right)}  \\
   {11S_{\parallel RR\phi }  - 4S_{\parallel \phi \phi \phi }  - S_{\parallel \phi zz} }  \\
   {11S_{\parallel RRz}  - S_{\parallel \phi \phi z}  - 4S_{\parallel zzz} }  \\
   {11S_{\parallel R\phi \phi }  - S_{\parallel Rzz}  - 4S_{\parallel RRR} }  \\
   {12S_{\parallel R\phi z} }  \\
   {11S_{\parallel Rzz}  - 4S_{\parallel RRR}  - S_{\parallel R\phi \phi } }  \\
   {4\left( {2S_{\parallel \phi \phi \phi }  - S_{\parallel RR\phi }  - S_{\parallel \phi zz} } \right)}  \\
   {11S_{\parallel \phi \phi z}  - S_{\parallel RRz}  - 4S_{\parallel zzz} }  \\
   {11S_{\parallel \phi zz}  - S_{\parallel RR\phi }  - 4S_{\parallel \phi \phi \phi } }  \\
   { - 4\left( {S_{\parallel RRz}  + S_{\parallel \phi \phi z}  - 2S_{\parallel zzz} } \right)}  \\

 \end{array} } \right)
\label{EQA07}
\end{equation}

From

\begin{equation}
{{\mathbf{\bar M}}_{\left[ 4 \right]}} = \frac{1}
{{4\pi }}\int_{S^2 }^{} {\bar M_{\left[ 4 \right]i,j} d\Omega },\label{EQA08}
\end{equation}

we get

	\begin{equation}
\left( {\begin{array}{*{20}c}
   {\bar M_{\left[ 4 \right]1,1} } &  \ldots  & {\bar M_{\left[ 4 \right]1,15} }  \\
    \vdots  &  \ddots  &  \vdots   \\
   {\bar M_{\left[ 4 \right]1,15} } &  \cdots  & {\bar M_{\left[ 4 \right]15,15} }  \\

 \end{array} } \right) = \frac{1}
{9} \cdot \left( {\begin{array}{*{20}c}
   1 & 0 & 0 & {\frac{1}
{2}} & 0 & {\frac{1}
{2}} & 0 & 0 & 0 & 0 & {\frac{3}
{{35}}} & 0 & {\frac{1}
{{10}}} & 0 & {\frac{3}
{{35}}}  \\
   0 & {\frac{4}
{7}} & 0 & 0 & 0 & 0 & {\frac{{12}}
{{35}}} & 0 & {\frac{8}
{{35}}} & 0 & 0 & 0 & 0 & 0 & 0  \\
   0 & 0 & {\frac{4}
{7}} & 0 & 0 & 0 & 0 & {\frac{8}
{{35}}} & 0 & {\frac{{12}}
{{35}}} & 0 & 0 & 0 & 0 & 0  \\
   {\frac{1}
{2}} & 0 & 0 & {\frac{{18}}
{{35}}} & 0 & {\frac{6}
{{35}}} & 0 & 0 & 0 & 0 & {\frac{1}
{2}} & 0 & {\frac{6}
{{35}}} & 0 & {\frac{1}
{{10}}}  \\
   0 & 0 & 0 & 0 & {\frac{{12}}
{{35}}} & 0 & 0 & 0 & 0 & 0 & 0 & {\frac{8}
{{35}}} & 0 & {\frac{8}
{{35}}} & 0  \\
   {\frac{1}
{2}} & 0 & 0 & {\frac{6}
{{35}}} & 0 & {\frac{{18}}
{{35}}} & 0 & 0 & 0 & 0 & {\frac{1}
{{10}}} & 0 & {\frac{6}
{{35}}} & 0 & {\frac{1}
{2}}  \\
   0 & {\frac{{12}}
{{35}}} & 0 & 0 & 0 & 0 & {\frac{4}
{7}} & 0 & {\frac{8}
{{35}}} & 0 & 0 & 0 & 0 & 0 & 0  \\
   0 & 0 & {\frac{8}
{{35}}} & 0 & 0 & 0 & 0 & {\frac{{12}}
{{35}}} & 0 & {\frac{8}
{{35}}} & 0 & 0 & 0 & 0 & 0  \\
   0 & {\frac{8}
{{35}}} & 0 & 0 & 0 & 0 & {\frac{8}
{{35}}} & 0 & {\frac{{12}}
{{35}}} & 0 & 0 & 0 & 0 & 0 & 0  \\
   0 & 0 & {\frac{{12}}
{{35}}} & 0 & 0 & 0 & 0 & {\frac{8}
{{35}}} & 0 & {\frac{4}
{7}} & 0 & 0 & 0 & 0 & 0  \\
   {\frac{3}
{{35}}} & 0 & 0 & {\frac{1}
{2}} & 0 & {\frac{1}
{{10}}} & 0 & 0 & 0 & 0 & 1 & 0 & {\frac{1}
{2}} & 0 & {\frac{3}
{{35}}}  \\
   0 & 0 & 0 & 0 & {\frac{8}
{{35}}} & 0 & 0 & 0 & 0 & 0 & 0 & {\frac{4}
{7}} & 0 & {\frac{{12}}
{{35}}} & 0  \\
   {\frac{1}
{{10}}} & 0 & 0 & {\frac{6}
{{35}}} & 0 & {\frac{6}
{{35}}} & 0 & 0 & 0 & 0 & {\frac{1}
{2}} & 0 & {\frac{{18}}
{{35}}} & 0 & {\frac{1}
{2}}  \\
   0 & 0 & 0 & 0 & {\frac{8}
{{35}}} & 0 & 0 & 0 & 0 & 0 & 0 & {\frac{{12}}
{{35}}} & 0 & {\frac{4}
{7}} & 0  \\
   {\frac{3}
{{35}}} & 0 & 0 & {\frac{1}
{{10}}} & 0 & {\frac{1}
{2}} & 0 & 0 & 0 & 0 & {\frac{3}
{{35}}} & 0 & {\frac{1}
{2}} & 0 & 1  \\

 \end{array} } \right),
\label{EQA09}
\end{equation}

and from its inverse

\[
{\mathbf{T}} \equiv \left( {T_{RRRR} ,T_{RRR\phi} ,T_{RRRz} ,T_{RR\phi \phi } ,T_{RR\phi z} ,T_{RRzz} ,T_{R\phi \phi \phi } ,T_{R\phi \phi z} ,T_{R\phi zz} ,T_{Rzzz} ,T_{\phi \phi \phi \phi } ,T_{\phi \phi \phi z} ,T_{\phi \phi zz} ,T_{\phi zzz} ,T_{zzzz} } \right)^T  =
\]
\begin{equation}
\frac{{315}}
{2} \cdot \left( {\begin{array}{*{20}c}
   {\frac{1}
{{47423}}\left( {276T_{\parallel RRRR}  + 2471T_{\parallel RR\phi \phi }  + 2471T_{\parallel RRzz}  - 5778T_{\parallel \phi \phi \phi \phi }  + 9534T_{\parallel \phi \phi zz}  - 5778T_{\parallel zzzz} } \right)}  \\
   {\frac{1}
{{64}}\left( {11T_{\parallel RRRz}  - 5T_{\parallel R\phi \phi \phi }  - 4T_{\parallel R\phi zz} } \right)}  \\
   {\frac{1}
{{64}}\left( {11T_{\parallel RRRz}  - 4T_{\parallel R\phi \phi z}  - 5T_{\parallel Rzzz} } \right)}  \\
   {\frac{1}
{{47423}}\left( {2471T_{\parallel RRRR}  + 5624T_{\parallel RR\phi \phi }  - 10520T_{\parallel RRzz}  + 2471T_{\parallel \phi \phi \phi \phi }  - 10520T_{\parallel \phi \phi zz}  + 9534T_{\parallel zzzz} } \right)}  \\
   {\frac{1}
{{16}}\left( {4T_{\parallel RR\phi z}  - T_{\parallel \phi \phi \phi z}  - T_{\parallel \phi zzz} } \right)}  \\
   {\frac{1}
{{47423}}\left( {2471T_{\parallel RRRR}  - 10520T_{\parallel RR\phi \phi }  + 5624T_{\parallel RRzz}  + 9534T_{\parallel \phi \phi \phi \phi }  - 10520T_{\parallel \phi \phi zz}  + 2471T_{\parallel zzzz} } \right)}  \\
   {\frac{1}
{{64}}\left( { - 5T_{\parallel RRR\phi }  + 11T_{\parallel R\phi \phi \phi \phi }  - 4T_{\parallel R\phi zz} } \right)}  \\
   {\frac{1}
{{16}}\left( { - T_{\parallel RRRz}  + 4T_{\parallel R\phi \phi z}  - T_{\parallel Rzzz} } \right)}  \\
   {\frac{1}
{{16}}\left( { - T_{\parallel RRR\phi }  - T_{\parallel R\phi \phi \phi }  + 4T_{\parallel R\phi zz} } \right)}  \\
   {\frac{1}
{{64}}\left( { - 5T_{\parallel RRRz}  - 4T_{\parallel R\phi \phi z}  + 11T_{\parallel Rzzz} } \right)}  \\
   {\frac{1}
{{47423}}\left( { - 5778T_{\parallel RRRR}  + 2471T_{\parallel RR\phi \phi }  + 9534T_{\parallel RRzz}  + 276T_{\parallel \phi \phi \phi }  + 2471T_{\parallel \phi \phi zz}  - 5778T_{\parallel zzzz} } \right)}  \\
   {\frac{1}
{{64}}\left( { - 4T_{\parallel RR\phi z}  + 11T_{\parallel \phi \phi \phi z}  - 5T_{\parallel \phi zzz} } \right)}  \\
   {\frac{1}
{{47423}}\left( {9534T_{\parallel RRRR}  - 10520T_{\parallel RR\phi \phi }  - 10520T_{\parallel RRzz}  + 2471T_{\parallel \phi \phi \phi \phi }  + 5624T_{\parallel \phi \phi zz}  + 2471T_{\parallel zzzz} } \right)}  \\
   {\frac{1}
{{64}}\left( { - 4T_{\parallel RR\phi z}  - 5T_{\parallel \phi \phi \phi z}  + 11T_{\parallel \phi zzz} } \right)}  \\
   {\frac{1}
{{47423}}\left( { - 5778T_{\parallel RRRR}  + 9534T_{\parallel RR\phi \phi }  + 2471T_{\parallel RRzz}  - 5778T_{\parallel \phi \phi \phi \phi }  + 2471T_{\parallel \phi \phi zz}  + 276T_{\parallel zzzz} } \right)}  \\

 \end{array} } \right).
\label{EQA10}
\end{equation}

\section{The ${\mathbf{V}}$-cumulants mixture distribution}
To disentangle the first and second cumulants of the  thick disk alone, we need up to the fourth order cumulants of the mixture. The ${\mathbf{V}}$-cumulants mixture distribution of third and fourth
order are introduced here. These moments exhibit a symmetry along the
azimuthal velocity vector of the mixture. Hence they differ from what is
laid out in the Appendices B, C, and D of \citet{2004A&A...427..131C}. Also
the notation differs in order to be consistent with the notation used in
our study. The third cumulants from which we can easily compute the two
components of the normalized vector ${\mathbf{\hat d}}$ defined in Eqn.\
\eqref{EQq}, are

\begin{equation}\label{EQB01}
\begin{gathered}
  {\rm K}_{RRR}  = \kappa _{zzz}  - \hat{d}_3^3 \kappa _{\phi\phi\phi}  + 3\hat{d}_3^2 \kappa _{\phi\phi z}  - 3\hat{d}_3 \kappa _{\phi z z}=0,  \hfill \\
  {\rm K}_{RRz}  = \hat{d}_1 \kappa _{\phi z z}  - \kappa _{Rzz}  + \hat{d}_3 \left( { - \hat{d}_3 \kappa _{R\phi\phi}  + 2\kappa _{R\phi z}  + \hat{d}_1 \hat{d}_3 \kappa _{\phi\phi\phi}  - 2\hat{d}_1 \kappa _{\phi\phi z} } \right)=0, \hfill \\
  {\rm K}_{Rzz}  = \kappa _{RRz}  - \hat{d}_3 \kappa _{RR\phi}  + \hat{d}_1 \left( {2\hat{d}_3 \kappa _{R\phi\phi}  - 2\kappa _{R\phi z}  - \hat{d}_1 \hat{d}_3 \kappa _{\phi\phi\phi}  + \hat{d}_1 \kappa _{\phi\phi z} } \right)=0, \hfill \\
  {\rm K}_{zzz}  =  - \kappa _{RRR}  + \hat{d}_1 \left( {3\kappa _{RR\phi}  + \hat{d}_1 \left( { - 3\kappa _{R\phi\phi}  + \hat{d}_1 \kappa _{\phi\phi\phi} } \right)} \right)=0. \hfill \\
\end{gathered}
\end{equation}
Once the values of $\hat{d}_1 $ and $\hat{d}_3 $ are computed as SVD solution of the previous overdetermined system, the
remaining cumulants can be calculated via
\begin{equation}\label{EQB02}
\begin{gathered}
  {\rm K}_{RR\phi }  = \kappa _{\phi zz}  - 2\hat{d}_3 \kappa _{\phi \phi z}  + \hat{d}_3^2 \kappa _{\phi \phi \phi },  \hfill \\
  {\rm K}_{R\phi \phi }  = \kappa _{\phi \phi z}  - \hat{d}_3 \kappa _{\phi \phi \phi },  \hfill \\
  {\rm K}_{R\phi z}  = \hat{d}_3 \kappa _{R\phi \phi }  - \kappa _{R\phi z}  + \hat{d}_1 \left( {\kappa _{\phi \phi z}  - \hat{d}_3 \kappa _{\phi \phi \phi } } \right), \hfill \\
  {\rm K}_{\phi \phi \phi }  = \kappa _{\phi \phi \phi },  \hfill \\
  {\rm K}_{\phi \phi z}  = \hat{d}_1 \kappa _{\phi \phi \phi }  - \kappa _{R\phi \phi },  \hfill \\
  {\rm K}_{\phi zz}  = \kappa _{RR\phi }  + \hat{d}_1 \left( { - 2\kappa _{R\phi \phi }  + \hat{d}_1 \kappa _{\phi \phi \phi } } \right), \hfill \\
\end{gathered}
\end{equation}
The fourth order equation can be computed as
\begin{equation}\label{EQB03}
\begin{gathered}
  {\rm K}_{RRRR}  = \kappa _{RRRR}  + \hat{d}_3 \left( { - 4\kappa _{\phi zzz}  + \hat{d}_3 \left( {6\kappa _{\phi \phi zz}  - 4\hat{d}_3 \kappa _{\phi \phi \phi z}  + \hat{d}_3^2 \kappa _{\phi \phi \phi \phi } } \right)} \right), \hfill \\
  {\rm K}_{RRRz}  =  - \kappa _{Rzzz}  + \hat{d}_1 \kappa _{\phi zzz}  +  \hfill \\
  \mbox{ }\mbox{ }\mbox{ }\mbox{ }\mbox{ }\mbox{ } \hat{d}_3 \left( {3\kappa _{R\phi zz}  - 3\hat{d}_1 \kappa _{\phi \phi zz}  + \hat{d}_3 \left( { - 3\kappa _{R\phi \phi z}  + \hat{d}_3 \kappa _{R\phi \phi \phi }  + 3\hat{d}_1 \kappa _{\phi \phi \phi z}  - \hat{d}_1 \hat{d}_3 \kappa _{\phi \phi \phi \phi } } \right)} \right), \hfill \\
  {\rm K}_{RRzz}  = \kappa _{RRzz}  - 2\hat{d}_3 \kappa _{RR\phi z}  + \hat{d}_3^2 \kappa _{RR\phi \phi }  +  \hfill \\
  \mbox{ }\mbox{ }\mbox{ }\mbox{ }\mbox{ }\mbox{ }\hat{d}_1 \left( { - 2\kappa _{R\phi zz}  + \hat{d}_1 \kappa _{\phi \phi zz}  + \hat{d}_3 \left( {4\kappa _{R\phi \phi z}  - 2\hat{d}_3 \kappa _{R\phi \phi \phi }  - 2\hat{d}_1 \kappa _{\phi \phi \phi z}  + \hat{d}_1 \hat{d}_3 \kappa _{\phi \phi \phi \phi } } \right)} \right), \hfill \\
  {\rm K}_{Rzzz}  =  - \kappa _{RRRz}  + \hat{d}_3 \kappa _{RRR\phi }  +  \hfill \\
  \mbox{ }\mbox{ }\mbox{ }\mbox{ }\mbox{ }\mbox{ } \hat{d}_1 \left( {3\kappa _{RR\phi z}  - 3\hat{d}_3 \kappa _{RR\phi \phi }  + \hat{d}_1 \left( { - 3\kappa _{R\phi \phi z}  + 3\hat{d}_3 \kappa _{R\phi \phi \phi }  + \hat{d}_1 \kappa _{\phi \phi \phi z}  - \hat{d}_1 \hat{d}_3 \kappa _{\phi \phi \phi \phi } } \right)} \right), \hfill \\
  {\rm K}_{zzzz}  = \kappa _{RRRR}  + \hat{d}_1 \left( { - 4\kappa _{RRR\phi }  + \hat{d}_1 \left( {6\kappa _{RR\phi \phi }  - 4\hat{d}_1 \kappa _{R\phi \phi \phi }  + \hat{d}_1^2 \kappa _{\phi \phi \phi \phi } } \right)} \right), \hfill \\
  {\rm K}_{RRR\phi }  = \kappa _{\phi zzz}  - \hat{d}_3\left( {3\kappa _{\phi \phi zz}  + \hat{d}_3\left( { - 3\kappa _{\phi \phi \phi z}  + \hat{d}_3\kappa _{\phi \phi \phi \phi } } \right)} \right), \hfill \\
  {\rm K}_{RR\phi z}  =  - \kappa _{R\phi zz}  + \hat{d}_1 \kappa _{\phi \phi zz}  + \hat{d}_3 \left( {2\kappa _{R\phi \phi z}  - \hat{d}_3 \kappa _{R\phi \phi \phi }  - 2\hat{d}_1 \kappa _{\phi \phi \phi z}  + \hat{d}_1 \hat{d}_3 \kappa _{\phi \phi \phi \phi } } \right), \hfill \\
  {\rm K}_{R\phi zz}  = \kappa _{RR\phi z}  - \hat{d}_3 \kappa _{RR\phi \phi }  + \hat{d}_1 \left( { - 2\kappa _{R\phi \phi z}  + 2\hat{d}_3 \kappa _{R\phi \phi \phi }  + \hat{d}_1 \kappa _{\phi \phi \phi z}  - \hat{d}_1 \hat{d}_3 \kappa _{\phi \phi \phi \phi } } \right), \hfill \\
  {\rm K}_{\phi zzz}  =  - \kappa _{RRR\phi }  + \hat{d}_1 \left( {3\kappa _{RR\phi \phi }  + \hat{d}_1 \left( { - 3\kappa _{R\phi \phi \phi }  + \hat{d}_1 \kappa _{\phi \phi \phi \phi } } \right)} \right), \hfill \\
  {\rm K}_{RR\phi \phi }  = \kappa _{\phi \phi zz}  + \hat{d}_3 \left( { - 2\kappa _{\phi \phi \phi z}  + {\hat{d}}_3 \kappa _{\phi \phi \phi \phi } } \right), \hfill \\
  {\rm K}_{R\phi \phi z}  =  - \kappa _{R\phi \phi z}  + \hat{d}_3 \kappa _{R\phi \phi \phi }  + \hat{d}_1 \left( {\kappa _{\phi \phi \phi z}  - \hat{d}_3 \kappa _{\phi \phi \phi \phi } } \right), \hfill \\
  {\rm K}_{\phi \phi zz}  = \kappa _{RR\phi \phi }  + \hat{d}_1 \left( { - 2\kappa _{R\phi \phi \phi }  + \hat{d}_1 \kappa _{\phi \phi \phi \phi } } \right), \hfill \\
  {\rm K}_{R\phi \phi \phi }  = \kappa _{\phi \phi \phi z}  - \hat{d}_3 \kappa _{\phi \phi \phi \phi },  \hfill \\
  {\rm K}_{\phi \phi \phi z}  =  - \kappa _{R\phi \phi \phi }  + \hat{d}_1 \kappa _{\phi \phi \phi \phi },  \hfill \\
  {\rm K}_{\phi \phi \phi \phi }  = \kappa _{\phi \phi \phi \phi }.  \hfill \\
\end{gathered}
\end{equation}
Once we have the whole set of the cumulants we define the tensor
\begin{equation}\label{EQB04}
{\mathbf{C}} \equiv \frac{1}
{{\sqrt {q^2  + 4} }}\left( {{\bm{\sigma }}_{\text{thick}}  - {\bm{\sigma }}_{\text{thin}} } \right) - q{\mathbf{d}}^{ \otimes 2},
\end{equation}
where ${\mathbf{d}} = \hat{d}_2 {\mathbf{\hat d}}$. The constraining equations
derived in \citet{2004A&A...427..131C} can be reduced to the following set
of fourteen scalar relations
\begin{equation}\label{EQB05}
\begin{gathered}
  \frac{{3{\rm K}_{RR\phi }^2 }}
{{{\rm K}_{RRRR} }} - \hat{d}_2^2  = \frac{{3{\rm K}_{RR\phi } {\rm K}_{R\phi z} }}
{{{\rm K}_{RRRz} }} - \hat{d}_2^2  = \frac{{{\rm K}_{RR\phi } {\rm K}_{zz\phi } {\text{ + 2}}{\rm K}_{R\phi z}^2 }}
{{{\rm K}_{RRzz} }} - \hat{d}_2^2  = \frac{{3{\rm K}_{R\phi z} {\rm K}_{\phi zz} }}
{{{\rm K}_{Rzzz} }} - \hat{d}_2^2  =  \hfill \\
  \frac{{3{\rm K}_{\phi zz}^2 }}
{{{\rm K}_{zzzz} }} - \hat{d}_2^2  = \frac{{3{\rm K}_{RR\phi } {\rm K}_{R\phi \phi } }}
{{{\rm K}_{RRR\phi } }} - \hat{d}_2^2  = \frac{{{\rm K}_{RR\phi } {\rm K}_{z\phi \phi } {\text{ + 2}}{\rm K}_{R\phi z} {\rm K}_{R\phi \phi } }}
{{{\rm K}_{RR\phi z} }} - \hat{d}_2^2  =  \hfill \\
  \frac{{{\rm K}_{zz\phi } {\rm K}_{R\phi \phi } {\text{ + 2}}{\rm K}_{R\phi z} {\rm K}_{z\phi \phi } }}
{{{\rm K}_{R\phi zz} }} - \hat{d}_2^2  = \frac{{3{\rm K}_{zz\phi } {\rm K}_{\phi \phi z} }}
{{{\rm K}_{\phi zzz} }} - \hat{d}_2^2  = \frac{1}
{{{\rm K}_{RR\phi } }}\left( {{\rm K}_{RR\phi \phi }  - \frac{{2{\rm K}_{R\phi \phi }^2 }}
{{\hat{d}_2^2 }}} \right)\frac{{\hat{d}_2^3 }}
{{c_{\phi\phi} }} - \hat{d}_2^2  =  \hfill \\
  \frac{1}
{{{\rm K}_{R\phi z} }}\left( {{\rm K}_{Rz\phi \phi }  - \frac{{2{\rm K}_{R\phi \phi } {\rm K}_{\phi \phi z} }}
{{\hat d }_2^2}} \right)\frac{{\hat{d}_2^3 }}
{c_{\phi\phi}} - \hat{d}_2^2  = \frac{1}
{{{\rm K}_{zz\phi } }}\left( {{\rm K}_{zz\phi \phi }  - \frac{{2{\rm K}_{z\phi \phi }^3 }}
{{\hat d }_2^2}} \right)\frac{{\hat{d}_2^3 }}
{c_{\phi\phi}} - \hat{d}_2^2  =  \hfill \\
  \frac{{{\rm K}_{R\phi \phi \phi } }}
{{3{\rm K}_{R\phi \phi } }}\frac{{\hat{d}_2^3 }}
{c_{\phi\phi}} - \hat{d}_2^2  = \frac{{{\rm K}_{z\phi \phi \phi } }}
{{3{\rm K}_{z\phi \phi } }}\frac{{\hat{d}_2^3 }}
{c_{\phi\phi}} - \hat{d}_2^2,  \hfill \\
\end{gathered}
\end{equation}
that we solve in a least-squares sense with respect to the elements
$c_{\phi\phi} $ and $\hat{d}_2 $. The final step to calculate the desired
results for $q$ is to take the derived values for $c_{\phi\phi} $ and
$\hat{d}_2 $ and to work out the last constraining equation from the
relations
\begin{equation}\label{EQB06}
\begin{gathered}
  {\rm K}_{zzz}  = \kappa _{zzz}  = 3c_{\phi\phi} \hat{d}_2  + 2q\hat{d}_2^3,  \hfill \\
  {\rm K}_{zzzz}  = \kappa _{zzzz}  = 3c_{\phi\phi}^2  - 2\left( {q^2  + 1} \right)\hat{d}_2^4.  \hfill \\
\end{gathered}
\end{equation}

\section{Thick disk parameters without knowledge of photometric distances}
By testing the method developed in this paper on a completely synthetic catalogue created with the Padua Galaxy Model \citep[e.g.,][ and references therein]{2004ASPC..317..203V} we can test our ability to recover the correct results, to refine the method and to improve its performance.

\begin{figure*}
\sidecaption
\includegraphics[width=10cm]{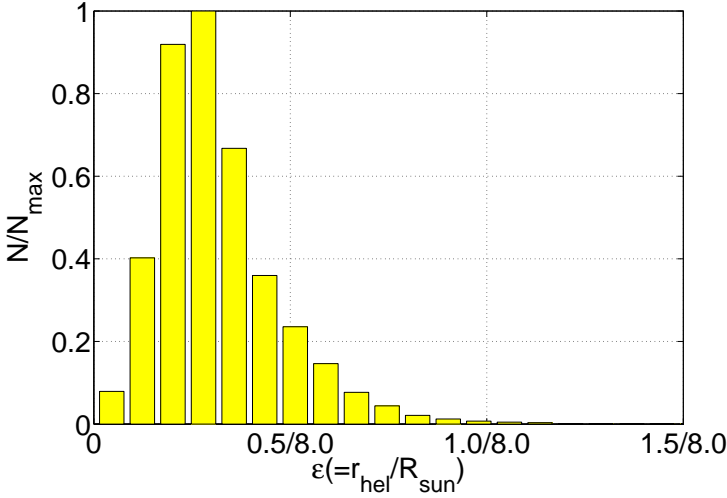}
\caption{Distance distribution for the selected sample of stars. The number is normalized to the highest value to evidence the peak position at 1. The $x$ axis shows distances divided by the adopted solar position (to better illustrate the parameter $\varepsilon$ for which the adopted approximations hold).}
\label{Figure06}
\end{figure*}

Moreover we obtain here a remarkable example of convergence of the method on the \textit{true RAVE data}, where the method is forced to work without the knowledge of the previously determined photometric distances by \cite{2010A&A...522A..54Z}.
A fundamental selection criterion in order to achieve this particular result is the cut in the surface gravity of the stars. In order to avoid the contamination by giant stars which can enter our sample because of their intrinsic luminosity despite their distances \citep[see, e.g.,][]{2008ApJ...685..261K, 2011ApJ...726..103K} we plot in Figure \ref{Figure06} the distance distribution of the dwarf stars selected with a cut in the surface gravity at $\log _{10} g = 3.5$.
Moreover in the introductory consideration (Section \ref{Preparingthedata}) we explained how the expansion over a parameter $\varepsilon$ of the radial component of the Galactic rotation leads to only a weak influence of the photometric distance errors on the results. This parameter has to be small, of the order of $\varepsilon  \simeq \frac{1}{8}$.
As evident by plotting the distances for an averaged sample of stars of the mixture ($ \approx 38,000$ stars) the distribution shows an $\varepsilon $ variation well within the $\frac{{1}} {8} \cong 0.12$ confirming that our selection cuts are able to retain stars with distances within a range of $r_{\rm{hel}}  \leqslant 1.0{\text{kpc}}$ without an a priori knowledge of the distance.

After that, technically speaking, this result is achieved by simply allowing the variation of the randomly sampled distances of each star not only within their photometrically determined errors, but along all of the lines of sight, working only with directions $(l,b)$ instead of the full parameter space of directions and distances $(l,b,d)$. The results are presented in Table \ref{Tabella04}. The results are remarkably similar to the ones presented in Table \ref{Tabella01} as expected from the selection criteria adopted in Eqn. \eqref{EQ06}.

\begin{table}
\caption{Values of the thick disk velocity dispersion tensor without the knowledge of the distances.}
 \centerline {\begin{tabular}{|ccc|rrr|}
 \hline
                    &                          &                        & $[{\mbox{ }\text{km s}}^{-1}]$ &   $[{\mbox{ }\text{km s}}^{-1}]$ & $[{\mbox{ }\text{km s}}^{-1}]$ \\
 \hline
 ${\sigma _{RR}}$ & ${\sigma _{R\phi}}$    & ${\sigma _{Rz}}$     & ${56.1 \pm 3.8}$ & ${29.4 \pm 17.2}$  & ${10.1 \pm 3.3}$ \\
                    & ${\sigma _{\phi\phi}}$ & ${\sigma _{\phi z}}$ &                   & ${46.1 \pm 6.7}$  & ${5.8 \pm 5.1}$ \\
                    &                          & ${\sigma _{zz}}$     &                   &                    & ${35.1 \pm 3.4}$ \\
\hline
\end{tabular}}
\label{Tabella04}
\end{table}

\end{document}